# A Detailed Analytical Derivation of the GN Model of Non-Linear Interference in Coherent Optical Transmission Systems

Pierluigi Poggiolini, Gabriella Bosco, Andrea Carena, Vittorio Curri, Yanchao Jiang
and Fabrizio Forghieri

*Abstract* — Recently, a perturbative model of non-linear fiber propagation in uncompensated optical transmission systems has been proposed, called GN-model [1]. Here, an extended and more detailed version of the GN-model derivation [1] is reported, providing deeper insight into the model. Some straightforward generalizations of the model are also proposed.

## I. INTRODUCTION

Recently, a perturbative model of non-linear fiber propagation in uncompensated optical transmission (UT) systems has been proposed [1], called the GN-model. For a general subject introduction and essential referencing of prior work on the topic of perturbative models, please see [1] and [4].

The model assumptions and derivation procedure were clearly outlined in [1]. The details of the derivation steps were however summarized, with only the key intermediate results reported. This paper has been written to provide an extended version of the GN-model derivation, which does not require the readers, who are interested in re-deriving the model themselves, to fill in the gaps between intermediate results on their own. By providing more details, substantial further insight is gained into some of the fundamental assumptions of the GN-model and how they play out to obtain the final results. This material could also be useful to those readers who are interested in further extending or modifying the model itself.

In addition, the key assumptions and derivation procedure of [1] is quite general as to the types of systems that it can address, but the results shown in [1] were restricted to links with identical spans. Here this limitation is removed.

The derivation presented here was carried out during the summer of 2010 and it was first reported in [3]. Its results were preliminarily published in [2] and then more extensively in [1]. Of late, two independent derivations of the GN-model [11], [12], still based on the key assumptions of [1], have been made public, confirming the correctness of the results of [1].

**Version 3** of this document provides a detailed model derivation. The level of detail is however non-homogeneous in all the parts of the derivation, as the process of transferring all the material from research results [3] into paper form is still ongoing. Later versions will take care of this aspect and provide further extensions.

**Version 4 and 5** of this document report some corrections on the formulas in Section IV-H, regarding non-identical spans.

**Version 6** corrects typos in Eqs. (103)-(104) and adds the derivation of the GN model formula for backward-pumped Raman amplification (Section IV-I).

**Version 7** corrects typos in Eqs. (103)-(104) and adds the derivation of the GN model formula for backward-pumped Raman amplification (Section IV-I).

**Version 8** adds closed-form analytical NLI formulas for the rather general case of EDFA-based systems with unequal spans and unequal channels, i.e., having different transmitted power per channel, different channel bandwidth and irregular spacing (Section V).

**Version 9** corrects a typo in formula (127).

**Version 10** makes it explicit that expressions (128) and (129) depend on the span index $n_s$, which was previously tacitly implied.

P. Poggiolini, G. Bosco, A. Carena, V. Curri and Y. Jiang are with Dipartimento di Elettronica e Telecomunicazioni, Politecnico di Torino, Corso Duca degli Abruzzi 24, 10129, Torino, Italy, e-mail: pierluigi.poggiolini@polito.it; F. Forghieri is with CISCO Photonics, Via Philips 12, 20052, Monza, Milano, Italy, fforghie@cisco.com. This work was supported by CISCO Systems within a sponsored research agreement (SRA) contract.

**Version 11** improves the description of dispersion, by replacing the simple $\beta_2$ parameter used in the previous versions with a generic $\beta(\omega)$. This level of generalization is kept till the end of Section IV-G. Then, a simplified model using both $\beta_2$ and $\beta_3$ is introduced. From Section IV-H onward, only $\beta_2$ is kept.

**Version 12** adds Appendix B where the derivation of Eq. (100) is now reported (arbitrary different spans). Sect. IV-I has also been upgraded so that now the distributed-gain case as well can handle spans that are different from one another.

**Version 13** (the current version) corrects a "sign" typo in Eq. (92) and (93), which did not have any consequence on any other formulas.

II. TABLE OF ACRONYMS AND SYMBOLS

- ASE: amplified spontaneous emission
- BER: bit error ratio
- CUT: channel under test
- D-FWM: degenerate FWM
- DM: dispersion management
- DSP: digital signal processing
- FWM: four-wave mixing
- GN-model: the non-linearity model dealt with in this paper, where "GN" stands for "Gaussian noise"
- ND-FWM: non degenerate FWM
- NLI: non-linear interference
- NLSE: non-linear Schroedinger equation
- OSNR: optical signal-to-noise-ratio
- PA: perturbative assumption
- PDL: polarization-dependent loss
- PM: polarization-multiplexed
- PRBS: pseudo-random binary sequence
- PSD: average power spectral density
- PWGN: periodic white Gaussian noise
- QPSK: quadrature phase-shift-keying
- QAM: quadrature amplitude polarization
- RP: random process
- RV: random variable
- Rx: receiver
- SNR: signal-to-noise ratio
- SPM: self-phase modulation
- Tx: transmitter
- UT: uncompensated transmission (no DM)
- XPM: cross-phase modulation

Some of the symbols used are:

- $\alpha$: optical field fiber loss [1/km], such that the optical **field** attenuates as $e^{-\alpha z}$; note that the optical **power** attenuates as $e^{-2\alpha z}$
- $\beta(f)$: propagation constant [1/km]
- $\beta_2$: second-order fiber dispersion [ps$^2$/km]
- $\beta_3$: third-order fiber dispersion [ps$^3$/km]
- $\gamma$: fiber non-linearity coefficient [1/(W km)]
- $R_s$: baud rate of an individual channel
- $T_s$: Tx symbol duration, such that $T_s = R_s^{-1}$

III. THE GENERALIZED SNR

Assuming linear propagation, assuming additive Gaussian ASE noise and neglecting PDL, the BER of any coherent system exploiting QAM modulation, including PM systems, can be expressed as a suitable function $\Psi$ of the SNR evaluated over the constellation scattering diagram, at the decision stage of the Rx, after DSP. Formally:

$$\text{BER} = \Psi(SNR) \quad (1)$$

The SNR is actually "measured" over the recovered signal "constellation" and can be written as:

$$\text{SNR} = \frac{\overline{A^2}}{\sigma_{ASE}^2} \quad (2)$$

where $\overline{A^2}$ is the average of the squared signal amplitude. If all symbols are equally likely, this quantity is just the sum of the squared distance of each noiseless constellation point from the origin, divided by the number of constellation points. Note that for PM systems the signal can be distributed over all four possible "quadratures", i.e., it is four-dimensional (4D). Therefore, in general, we have:

$$\overline{A^2} = \overline{A_{I,x}^2} + \overline{A_{Q,x}^2} + \overline{A_{I,y}^2} + \overline{A_{Q,y}^2} \quad (3)$$

where $I$, $Q$, stand for "in-phase" and "quadrature" and $x$, $y$ identify the two orthogonal polarizations. The symbol $\sigma_{ASE}^2$ represents the total variance of the 4D signal, due to ASE noise, that is:

$$\sigma_{ASE}^2 = \sigma_{I,x}^2 + \sigma_{Q,x}^2 + \sigma_{I,y}^2 + \sigma_{Q,y}^2 \quad (4)$$

Assuming that the Rx is essentially ideal, noise on each quadrature is statistically independent of the other quadratures and all quadrature variances are equal:

$$\sigma_{I,x}^2 = \sigma_{Q,x}^2 = \sigma_{I,y}^2 = \sigma_{Q,y}^2 \qquad (5)$$

The function $\Psi$ in Eq. (1) depends on the modulation format. As an example, for PM-QPSK, we have:

$$\mathrm{BER} = \frac{1}{2}\mathrm{erfc}\left(\sqrt{\mathrm{SNR}/2}\right) \qquad (6)$$

For other formats, other expressions apply. See Appendix A in [1] or any textbook on digital transmission theory, such as [5]. We have defined the SNR based on the Rx scattering diagram after DSP. However, under certain assumptions, the SNR can also be computed based on the optical signal impinging on the detectors. Ideally, if the only cause of signal degradation is ASE noise, then the noise variance on the constellation can be found as:

$$\sigma_{ASE}^2 = \int_{-\infty}^{\infty} G_{ASE} \cdot \left|H_{Rx}(f)\right|^2 df \qquad (7)$$

where $G_{ASE}$ is the power spectral density (PSD) of the dual-polarization ASE noise and $H_{Rx}(f)$ is the overall baseband scalar transfer function of the coherent Rx, including an equalizer if present. Note that we assume $G_{ASE}$ to be a *unilateral* noise PSD, then translated to baseband. To avoid any ambiguity in the definitions, we point out that in this notation $G_{ASE}$ for a single EDFA amounts to:

$$G_{ASE} = F(G-1)h\nu_{CUT} \qquad (8)$$

where $G$ is the gain of the amplifier and $F = 2n_{sp}$ is the noise figure, with $n_{sp} \geq 1$ the noise enhancement factor.

As for the signal factor $\overline{A^2}$ in Eq. (2), its value depends on various details regarding the transmitted waveforms and the actual Rx baseband transfer function $H_{Rx}(f)$. However, assuming that the Tx signal does not suffer from ISI and $H_{Rx}(f)$ is shaped so that *matched* filtering occurs, it can be shown that:

$$\overline{A^2} = P_{Rx} \cdot R_s^{-1} \int_{-\infty}^{\infty} \left|H_{Rx}(f)\right|^2 d\nu \qquad (9)$$

Note that under the same assumptions it also turns out that:

$$\mathrm{SNR} = \frac{\overline{A^2}}{\sigma_{ASE}^2} = \frac{P_{Rx}}{G_{ASE}R_s} = \frac{P_{Rx}}{G_{ASE}B_N}\frac{B_N}{R_s} = \\ = \frac{P_{Rx}}{P_{ASE,B_N}}\frac{B_N}{R_s} = \frac{B_N}{R_s}\mathrm{OSNR}_{B_N} \qquad (10)$$

where $\mathrm{OSNR}_{B_N}$ is the optical signal-to-noise-ratio over a bandwidth $B_N$. However, under more general conditions, such as non-matched filtering, the relationship between OSNR and SNR is more complex. In this paper we will always assume that $H_{Rx}(f)$ implements matched filtering. This assumption is not unrealistic since the typical adaptive equalizers present in the Rx DSPs tend to converge to matched filtering.

One of the fundamental assumptions on which the GN model is based is that the effect of NLI on WDM signals can be modeled as *additive Gaussian noise*, statistically independent of ASE noise [1]. The direct consequence of such assumption is that ASE and NLI noise contributions simply add up in variance. The BER of the CUT still depends on SNR through Eq. (1), but the definition of SNR needs to be modified to include NLI noise:

$$\mathrm{SNR} = \frac{\overline{A^2}}{\sigma_{ASE}^2 + \sigma_{NLI}^2} \qquad (11)$$

with:

$$\sigma_{NLI}^2 = \int_{-\infty}^{\infty} G_{NLI}(f)\left|H_{Rx}(f)\right|^2 df \qquad (12)$$

where $G_{NLI}(f)$ is the unilateral PSD of NLI, down-converted to baseband. The fundamental quantity that needs to be assessed to characterize the non-linear system behavior is therefore $G_{NLI}(f)$.

IV. THE NLI MODEL DERIVATION

In the following we concentrate on the derivation of $G_{NLI}(f)$. We start out by providing a suitable transmitted signal model.

*A. The Signal Model*

The signal model was introduced and extensively discussed and justified in [1]. Therefore, here we only recall its main features and the defining formulas. The interested reader should see Section II-B and Appendix B in [1]. The signal model is one of the key aspects enabling the derivation of the GN-model.

We call $E(t)$ the WDM Tx signal in time-domain. It is a RP generated by the transmission of random independent symbols on each WDM channel. As it propagates along the fiber it takes on a $z$-dependence as well, where $z$ is the distance coordinate along the link. We then denote the propagating signal with the symbol $E(z,t)$, with: $E(0,t) = E(t)$.

As discussed in [1], dispersion causes the signal of each individual channel, and as a result the overall WDM signal $E(z,t)$, to take on a complex zero-mean Gaussian statistical distribution, whose variance is solely determined by the average power of the signal itself. Therefore, the signal model must be a Gaussian RP. In addition, the signal model is chosen so that it consists of spectral lines, to make it possible to tackle the non-linear propagation problem using methods that address spectral lines. This is easily obtained by assuming the signal model to be periodic. Such periodicity assumption actually causes no loss of generality since the period can be assumed arbitrarily long. At any rate, later on this constraint will actually be lifted by letting the period tend to infinity.

Finally, the average PSD of the signal model must be shaped as that of an actual transmitted signal, which we call $G_{Tx}(f)$. Since the signal model is periodic, this constraint means that the coefficients of the spectral lines vary according to $G_{Tx}(f)$.

In detail, we adopt an overall WDM signal model satisfying the following constraints 1-3:
1. it is a zero-mean complex Gaussian RP with uncorrelated phase and quadrature components
2. it is periodic of period $T_0$, where $T_0$ is an integer multiple $W$ of the symbol duration $T_s$
3. its average PSD is shaped according to that of an actual WDM signal $G_{Tx}(f)$.

To satisfy 1-3 above, the signal model $E(t)$ is based on a filtered complex periodic white Gaussian noise (PWGN) process, as follows. First, we consider a PWGN process of period $T_0$, which can be written in frequency-domain using the Karhunen-Loeve expansion [5] as:

$$\text{PWGN}(f) = \sqrt{f_0} \sum_{n=\infty}^{+\infty} \xi_n \delta(f - nf_0) \quad (13)$$

where the $\xi_n$'s are complex Gaussian RVs of unit variance, independent of one another. Note that the average PSD of such RP is made up of identical-power spectral lines:

$$G_{\text{PWGN}}(f) = f_0 \sum_{n=-\infty}^{+\infty} \mathbf{E}\{|\xi_n|^2\} \delta(f - nf_0)$$
$$= f_0 \sum_{n=-\infty}^{+\infty} \delta(f - nf_0) \quad (14)$$

where $\mathbf{E}$ is the statistical average (or expectation) operator. Then, we ideally filter the PWGN of Eq. (13) through a transfer function:

$$H(f) = \sqrt{G_{Tx}(f)} \quad (15)$$

and obtain the wanted signal model, in frequency-domain:

$$E(f) = \text{PWGN}(f) \cdot H(f)$$
$$= \sqrt{f_0 G_{Tx}(f)} \sum_{n=-\infty}^{+\infty} \xi_n \delta(f - nf_0) \quad (16)$$

In time-domain:

$$E(t) = \sqrt{f_0 G_{Tx}(nf_0)} \sum_{n=-\infty}^{+\infty} \xi_n e^{j2\pi nf_0 t} \quad (17)$$

This signal model certainly satisfies constraints 1-2 because it is obtained from a process (13) that satisfies both, and the linear filtering of Eq. (16) does not alter such features. As for constraint 3, we point out that the average PSD of the process $E(t)$ is:

$$G_E(f) = f_0 G_{Tx}(f) \sum_{n=-\infty}^{+\infty} \mathbf{E}\{|\xi_n|^2\} \delta(f - nf_0)$$
$$= G_{Tx}(f) f_0 \sum_{n=-\infty}^{+\infty} \delta(f - nf_0) \quad (18)$$

Such PSD has a spectral "envelope" modeled after that of the WDM signal $G_{Tx}(f)$. As a result, 3 is satisfied too.

Having established the signal model as Eq. (17), the question arises of how accurately it models an actually transmitted signal. This aspect is thoroughly discussed in [1], Appendix B, and is omitted here. The result is that indeed (17) is an accurate signal model in the context of UT.

In the remainder of this section, we discuss signal power, as this aspect will be important later on. The average model signal power $P_E$ can be found by integrating its average PSD. Using (18) we have:

$$P_E = \int_{-\infty}^{\infty} G_E(f) df = f_0 \sum_{n=-\infty}^{+\infty} G_{Tx}(nf_0) \quad (19)$$

We then remark that, as a direct consequence of constraint 3, it must also be:

$$P_E = P_{Tx} \quad (20)$$

where $P_{Tx}$ is the actually transmitted average signal power, which amounts to:

$$P_{Tx} = \int_{-\infty}^{\infty} G_{Tx}(f) df \quad (21)$$

If we compare the right-hand sides of (19) and (21) we see that (20) is at least approximately met, because:

$$P_E = f_0 \sum_{n=-\infty}^{+\infty} G_{Tx}(nf_0) \approx \int_{-\infty}^{\infty} G_{Tx}(f)df = P_{Tx} \quad (22)$$

The above approximate equality is asymptotically exact as $T_0 \to \infty$ and $f_0 \to 0$. Since $T_0$ can be assumed arbitrarily large, (20) is hence verified to any arbitrary accuracy. Therefore, in the following we will simply assume:

$$P_E = P_{Tx} \quad (23)$$

A final important point regarding signal power is that $P_E$, as defined in (19), is technically a statistical *ensemble average* of the power of $E(t)$. The power of each individual model signal instance (or "realization") is instead given by:

$$\tilde{P}_E = f_0 \sum_{n=-\infty}^{+\infty} G_{Tx}(nf_0)|\xi_n|^2 \quad (24)$$

Note that $\tilde{P}_E$ is a RV itself, with $P_E = \mathbf{E}\{\tilde{P}_E\}$. However, for increasing $T_0$ the number of contributions to the summation in (24) grows and by simple rules of statistical convergence it can be concluded that the variance of $\tilde{P}_E$ decreases as $T_0$ is increased. Hence, at least RMS, $\tilde{P}_E \to P_E$ as $T_0$ is increased. Since $T_0$ can be chosen arbitrarily large, such convergence can be made arbitrarily accurate. In the following we will therefore simply assume that the power of each signal instance is a constant too, i.e., we shall assume:

$$\tilde{P}_E = P_E = P_{Tx}. \quad (25)$$

*B. The NLSE*

The NLSE is the fundamental equation of single-polarization fiber non-linear dispersive propagation. We first discuss the GN-model derivation in the context of single-polarization, for convenience, and then generalize to dual-polarization. We employ the general propagation constant $\beta$, which varies as a function of frequency. We then assume for now that the fiber parameters $\alpha$ and $\beta$ do not vary along the fiber. This assumption can be lifted as well and is taken here for convenience.
The NLSE can then be written in time-domain as:

$$\frac{\partial}{\partial z}E(z,t) = -j\beta(t)E(z,t) - \alpha E(z,t) + \\ -j\gamma E(z,t)E^*(z,t)E(z,t) \quad (26)$$

where the last term is due to the Kerr effect in the fiber. Applying a Fourier transformation to both sides, the frequency-domain NLSE is obtained:

$$\frac{\partial}{\partial z}E(z,f) = -j\beta(f)E(z,f) - \alpha E(z,f) + \\ -j\gamma E(z,f) * E^*(z,-f) * E(z,f) \quad (27)$$

where the symbol "$*$" stands for "convolution product". We elect to work with the frequency-domain NLSE because it reduces the NLSE to an ordinary differential equation rather than a partial differential equation, by eliminating the time-derivative. We now re-write it as:

$$\frac{\partial}{\partial z}E(z,f) = -j\beta(f)E(z,f) - \alpha E(z,f) + Q_{NLI}(z,f) \quad (28)$$

where:

$$Q_{NLI}(z,f) = -j\gamma E(z,f) * E^*(z,-f) * E(z,f) \quad (29)$$

is the Kerr term.
Before proceeding, we investigate the nature of $Q_{NLI}(z,f)$ and we first do so at the fiber input, that is at $z=0$, where we can write:

$$Q_{NLI}(0,f) = -j\gamma E(0,f) * E^*(0,-f) * E(0,f) \\ = -j\gamma \int_{-\infty}^{\infty} \int_{-\infty}^{\infty} E(0,f_1)E^*(0,f_1-f_2)E(0,f-f_2)df_1df_2 \quad (30)$$

We then substitute the signal model of Eq. (16) into Eq. (30):

$$Q_{NLI}(0,f) = -j\gamma f_0^{\frac{3}{2}} \sum_{m=-\infty}^{\infty} \sum_{n=-\infty}^{\infty} \sum_{k=-\infty}^{\infty} \xi_m \xi_n^* \xi_k \\ \int_{-\infty}^{\infty} \int_{-\infty}^{\infty} \sqrt{G_{Tx}(f_1)G_{Tx}(f_1-f_2)G_{Tx}(f-f_2)} \\ \delta(f_1-mf_0)\delta(f_1-f_2-nf_0)\delta(f-f_2-kf_0)df_2df_1 \quad (31)$$

Solving for the integrals is straightforward, thanks to the properties of the deltas. As a result:

$$Q_{NLI}(0,f) = -j\gamma f_0^{\frac{3}{2}} \sum_{m=-\infty}^{\infty} \sum_{n=-\infty}^{\infty} \sum_{k=-\infty}^{\infty} \\ \sqrt{G_{Tx}(mf_0)G_{Tx}(nf_0)G_{Tx}(kf_0)} \xi_m \xi_n^* \xi_k \delta(f-[m-n+k]f_0) \quad (32)$$

Note that the summations in Eq. (32) formally extend from minus to plus infinity but, in actuality, the shaping spectrum

$G_{Tx}(if_0)$ is non-zero only over a finite range of frequencies so that the number of-non-zero terms in Eq. (32) is finite. Not all the terms in Eq. (32) are homogenous. Some have special features and should be dealt with separately. First, we remark that Eq. (32) could be equivalently re-written as:

$$Q_{NLI}(0,f) = -j\gamma f_0^{\frac{3}{2}} \sum_{i=-\infty}^{\infty} \delta(f - if_0) \cdot \sum_{m,n,k \in A_i} \xi_m \xi_n^* \xi_k \sqrt{G_{Tx}(mf_0)G_{Tx}(nf_0)G_{Tx}(kf_0)} \quad (33)$$

where $A_i$ is the set of all triples $(m,n,k)$ such that $m - n + k = i$, that is:

$$A_i \equiv \{(m,n,k) : m - n + k = i\} \quad (34)$$

Among all the possible triples $(m,n,k)$ in $A_i$ we identify a subset $X_i$ for which $[m = n \text{ or } k = n]$, that is:

$$X_i \equiv \{(m,n,k) : [m - n + k = i] \text{ and } [m = n \text{ or } k = n]\} \quad (35)$$

We then define the coset $\tilde{A}_i$ as:

$$\tilde{A}_i = A_i - X_i \quad (36)$$

We can consequently decompose $Q_{NLI}(0,f)$ into two separate contributions:

$$Q_{NLI}(0,f) = Q_{NLI,\tilde{A}_i}(0,f) + Q_{NLI,X_i}(0,f) \quad (37)$$

where the two terms in the right-hand side are similar to Eq. (33) except the inner summations are executed only on $\tilde{A}_i$ and $X_i$, respectively:

$$Q_{NLI,\tilde{A}_i}(0,f) = -j\gamma f_0^{\frac{3}{2}} \sum_{i=-\infty}^{\infty} \delta(f - if_0) \cdot \sum_{m,n,k \in \tilde{A}_i} \xi_m \xi_n^* \xi_k \sqrt{G_{Tx}(mf_0)G_{Tx}(nf_0)G_{Tx}(kf_0)} \quad (38)$$

$$Q_{NLI,X_i}(0,f) = -j\gamma f_0^{\frac{3}{2}} \sum_{i=-\infty}^{\infty} \delta(f - if_0) \cdot \sum_{m,n,k \in X_i} \xi_m \xi_n^* \xi_k \sqrt{G_{Tx}(mf_0)G_{Tx}(nf_0)G_{Tx}(kf_0)} \quad (39)$$

By using Eq. (35) we can rewrite $Q_{NLI,X_i}(0,f)$, finding:

$$Q_{NLI,X_i}(0,f) = -j2\gamma f_0^{\frac{3}{2}} \sum_{i=-\infty}^{\infty} \delta(f - if_0) \sqrt{G_{Tx}(if_0)} \xi_i \cdot \sum_n G_{Tx}(nf_0)|\xi_n|^2 \quad (40)$$

The last summation in (40) is $\tilde{P}_E$ as defined in Eq. (24). Recalling assumption (25), we can write:

$$Q_{NLI,X_i}(0,f) = -j2\gamma \sqrt{f_0} P_{Tx} \cdot \sum_{i=-\infty}^{\infty} \sqrt{G_{Tx}(if_0)} \xi_i \delta(f - if_0) = -j2\gamma P_{Tx} E(f) \quad (41)$$

This result shows that the effect of $Q_{NLI,X_i}(0,f)$ on the NLSE is that of adding a constant-coefficient multiplying the unknown $E(t)$. At the fiber input, therefore, the NLSE can be rewritten as:

$$\left.\frac{\partial}{\partial z} E(z,f)\right|_{z=0} = [-j\beta(f) - j2\gamma P_{TX} - \alpha] E(0,f) + Q_{NLI,\tilde{A}_i}(0,f) \quad (42)$$

The question is then what form $Q_{NLI,X_i}(0,f)$ takes on at a generic distance $z > 0$. It can be shown that it remains very similar to Eq. (41), namely:

$$Q_{NLI,X_i}(z,f) = -j2\gamma P_{Tx}(z) \cdot E(z,f) \quad (43)$$

where $P_{Tx}(z)$ is the total signal power in fiber at $z > 0$. A discussion and proof of Eq. (43) is reported in Appendix A. To evaluate $P_{Tx}(z)$ we first recall the well-known result that both dispersion and the overall non-linear Kerr term in (28) are power-preserving terms: they cannot cause either extra loss or gain, but only a re-distribution of power among different frequencies. Therefore:

$$P_{Tx}(z) = P_{Tx} \cdot e^{-2\alpha z} \quad (44)$$

As a result, the NLSE at any distance $z$ becomes:

$$\frac{\partial}{\partial z} E(z,f) = \left[-j\beta(f) - j2\gamma P_{TX} e^{-2\alpha z} - \alpha\right] E(z,f) + Q_{NLI,\tilde{A}_i}(z,f) \quad (45)$$

We now introduce a fundamental assumption on which the model is based: we assume that the Kerr term $Q_{NLI,\tilde{A}_i}(z,f)$ acts as a pure source term in (42), i.e., it acts as if it was independent of the equation unknown $E(z,f)$. If so, according to standard results, a fully analytical solution to (42) can be written as:

$$E(z,f) = e^{\Gamma(f,z)} \cdot \int_0^z e^{-\Gamma(f,z')} Q_{NLI,\tilde{A}_i}(z',f)\, dz' + e^{\Gamma(f,z)} E(0,f) \quad (46)$$

where:

$$\Gamma(z,f) = \int_0^z \left[ -j\beta(f) - j2\gamma P_{TX} e^{-2\alpha\zeta} - \alpha \right] d\zeta \quad (47)$$
$$= -j\beta(f)z - j2\gamma P_{TX} z_{eff}(z) - \alpha z$$

with:

$$z_{eff}(z) = \frac{1 - e^{-2\alpha z}}{2\alpha} \quad (48)$$

where $z_{eff}$ is the so-called "effective length", related to the physical length $z$.
From Eq. (46), we can write:

$$E(z,f) \approx E_{LIN}(z,f) + E_{NLI}(z,f) \quad (49)$$

where:

$$E_{LIN}(z,f) = e^{\Gamma(f,z)} E(0,f) \quad (50)$$

is the "linear" solution, i.e., what one gets in the absence of non-linearity, and:

$$E_{NLI}(z,f) = e^{\Gamma(f,z)} \cdot \int_0^z e^{-\Gamma(f,z')} Q_{NLI,\tilde{A}_i}(z',f)\, dz' \quad (51)$$

is the *"non-linear interference"*, or NLI. Finding the PSD of $E_{NLI}(z,f)$ is the main goal of the whole model derivation and therefore (51) is a key result.

The assumption of the independence of $Q_{NLI,\tilde{A}_i}(z,f)$ on $E(z,f)$, which makes it possible to write (51), is clearly an approximation, since $Q_{NLI,\tilde{A}_i}(z,f)$ is a function of $E(z,f)$. However, the double convolution in (29) thoroughly scrambles the signal frequency components contributing to $Q_{NLI,\tilde{A}_i}(z,f)$, suggesting that $Q_{NLI,\tilde{A}_i}(z,f)$ at a certain frequency $f = f_0$ can indeed be thought of as being independent of $E(z,f)$ at the same frequency $f = f_0$. In this version of this document, such assumption is just considered as one of the approximations leading to the GN model, whose validity has been confirmed by the overall validation tests conducted elsewhere [1], [2]. It is however likely that this assumption can be justified in more formal terms. Such formal justification could be included in later versions of this document.

Even assuming that $Q_{NLI,\tilde{A}_i}(z,f)$ can be considered an independent source term, there remains the problem of actually computing it. To compute $Q_{NLI,\tilde{A}_i}(z,f)$, one must resort to Eq. (29), which however requires the prior knowledge of $E(z,f)$, that is the unknown quantity that we would like to obtain based on initial conditions and on $Q_{NLI,\tilde{A}_i}(z,f)$ itself.

To overcome this problem, we make a further key assumption. We assume that the regime of interest for communications systems is one of low-to-moderate non-linearity. In other words, we assume that the linear solution $E_{LIN}(z,f)$ of Eq. (50), is not too different from $E(z,f)$.

This assumption amounts to a *perturbative* approach to the solution of the NLSE, which can only be accurate as long as the perturbation remains "small". The validity of such assumption has been verified a posteriori by the successful model validations in [1] and [2]. It can also be justified by formally showing that the optimum operating regime of optical systems is one where non-linearity disturbance is always less than ASE noise-induced disturbance (see [4], Sect. XII). This finding has been confirmed both simulatively [6] and experimentally [7]. This clearly means that the signal constellation is not completely disrupted, but only *perturbed* by NLI, so that a *perturbative* approach like the one we propose appears reasonable.

Note that a similar approach is also commonplace for the analytical assessment of FWM due to unmodulated carriers and is known in that context as the "undepleted pump assumption". In this context we prefer to call it *perturbative assumption* (PA). According to all the previous assumptions, we can then write:

$$Q_{NLI}(z,f) \approx -j\gamma E_{LNI}(z,f) * E^*_{LNI}(z,-f) * E_{LNI}(z,f) \quad (52)$$

Substituting (50) into (52):

$$Q_{NLI}(z,f) = -j\gamma e^{-3\alpha z} e^{-j2\gamma P_{Tx} z_{eff}(z)} \int_{-\infty}^{\infty} \int_{-\infty}^{\infty} e^{-j\beta(f_1)z} E(0,f_1) \cdot$$
$$e^{j\beta(f_1 - f_2)z} E^*(0, f_1 - f_2) e^{-j\beta(f - f_2)z} E(0, f - f_2) df_1 df_2 = \quad (53)$$
$$= -j\gamma e^{-j2\gamma P_{Tx} z_{eff}(z)} e^{-3\alpha z} \int_{-\infty}^{\infty} \int_{-\infty}^{\infty} e^{-j[\beta(f_1) - \beta(f_1 - f_2) + \beta(f - f_2)]z} \cdot$$
$$\cdot E(0,f_1) E^*(0, f_1 - f_2) E(0, f - f_2) df_1 df_2$$

We then substitute the signal model of Eq. (16) into Eq. (53):

$$Q_{NLI}(z,f) = -j\gamma f_0^{\frac{3}{2}} e^{-3\alpha z} e^{-j2\gamma P_{Tx} z_{eff}(z)} \sum_{m=-\infty}^{\infty} \sum_{n=-\infty}^{\infty} \sum_{k=-\infty}^{\infty} \xi_m \xi_n^* \xi_k$$
$$\int_{-\infty}^{\infty} \int_{-\infty}^{\infty} \sqrt{G_{Tx}(f_1) G_{Tx}(f_1-f_2) G_{Tx}(f-f_2)} e^{-j[\beta(f_1)-\beta(f_1-f_2)+\beta(f-f_2)]z}$$
$$\delta(f_1 - mf_0)\delta(f_1 - f_2 - nf_0)\delta(f - f_2 - kf_0) df_2 df_1 \quad (54)$$

Solving for the integrals, one finds:

$$Q_{NLI}(z,f) = -j\gamma f_0^{\frac{3}{2}} e^{-3\alpha z} e^{-j\gamma 2 P_{Tx} z_{eff}(z)} \sum_{m=-\infty}^{\infty} \sum_{n=-\infty}^{\infty} \sum_{k=-\infty}^{\infty}$$
$$\sqrt{G_{Tx}(mf_0) G_{Tx}(nf_0) G_{Tx}(kf_0)} e^{-j[\beta(mf_0)-\beta(nf_0)+\beta(kf_0)]z} \quad (55)$$
$$\xi_m \xi_n^* \xi_k \delta(f - [m-n+k]f_0)$$

which can be rewritten as:

$$Q_{NLI}(z,f) = -j\gamma f_0^{\frac{3}{2}} e^{-3\alpha z} e^{-j2\gamma P_{Tx} z_{eff}(z)}$$
$$\sum_{i=-\infty}^{\infty} \delta(f - if_0) \sum_{m,n,k \in A_i} \sqrt{G_{Tx}(mf_0) G_{Tx}(nf_0) G_{Tx}(kf_0)}$$
$$\xi_m \xi_n^* \xi_k e^{-j[\beta(mf_0)-\beta(nf_0)+\beta(kf_0)]z} \quad (56)$$

We now restrict the summation to only the terms in $\tilde{A}_i$, because we have already dealt with the terms belonging to $X_i = (A_i - \tilde{A}_i)$, whose effect (43) is already taken into account by the term $-j2\gamma P_{Tx} z_{eff}(z)$ in Eq. (47). We finally obtain:

$$Q_{NLI,\tilde{A}_i}(z,f) = -j\gamma f_0^{\frac{3}{2}} e^{-3\alpha z} e^{-j2\gamma P_{Tx} z_{eff}(z)}$$
$$\sum_{i=-\infty}^{\infty} \delta(f - if_0) \sum_{m,n,k \in \tilde{A}_i} \sqrt{G_{Tx}(mf_0) G_{Tx}(nf_0) G_{Tx}(kf_0)}$$
$$\xi_m \xi_n^* \xi_k e^{-j[\beta(mf_0)-\beta(nf_0)+\beta(kf_0)]z} \quad (57)$$

Therefore, we have obtained the proper source term that, substituted into (51), makes it possible to compute an approximate expression of $E_{NLI}(z,f)$.

*C. The NLI Field*

We can now compute the NLI field $E_{NLI}(z,f)$. We do this first for one span, and then derive the solution after any number of spans. Inserting Eq. (57) into Eq. (51), we get:

$$E_{NLI}(z,f) = e^{-j\beta(f)z} e^{-j2\gamma P_{Tx} z_{eff}(z)} e^{-\alpha \cdot z} \cdot$$
$$\int_0^z e^{j\beta(f)z'} e^{j2\gamma P_{Tx} z_{eff}(z')} e^{\alpha z'} Q_{NLI}(z',f) dz' =$$
$$= -j\gamma f_0^{\frac{3}{2}} e^{-j\beta(f)z} e^{-j2\gamma P_{Tx} z_{eff}(z)} e^{-\alpha \cdot z}$$
$$\sum_{i=-\infty}^{\infty} \delta(f - if_0) \sum_{m,n,k \in \tilde{A}_i} \sqrt{G_{Tx}(mf_0) G_{Tx}(nf_0) G_{Tx}(kf_0)}$$
$$\xi_m \xi_n^* \xi_k \cdot \int_0^z e^{-2\alpha z'} e^{j\{\beta([m-n+k]f_0)-\beta(mf_0)+\beta(nf_0)-\beta(kf_0)\}z'} dz'$$
$$\quad (58)$$

The last integral is a well-known one, akin to the FWM efficiency. Solving for it, we get:

$$E_{NLI}(z,f) = \sum_{i=-\infty}^{\infty} \delta(f - if_0) \cdot \left[ -j\gamma f_0^{\frac{3}{2}} e^{-j\beta(if_0)z} e^{-j2\gamma P_{Tx} z_{eff}(z)} \right.$$
$$e^{-\alpha z} \sum_{m,n,k \in \tilde{A}_i} \sqrt{G_{Tx}(mf_0) G_{Tx}(nf_0) G_{Tx}(kf_0)} \xi_m \xi_n^* \xi_k$$
$$\left. \frac{1 - e^{-2\alpha z} e^{j\{\beta([m-n+k]f_0)-\beta(mf_0)+\beta(nf_0)-\beta(kf_0)\}z}}{2\alpha - j\{\beta([m-n+k]f_0) - \beta(mf_0) + \beta(nf_0) - \beta(kf_0)\}} \right]$$

$$(59)$$

We now remark that Eq. (59) has the overall form:

$$E_{NLI}(z,f) = \sum_i \mu_i \delta(f - if_0) \quad (60)$$

that is the NLI disturbance Fourier transform is a set of deltas. This means that $E_{NLI}(z,f)$ is still a periodic signal of period $T_0$. According to the theory of periodic signals, the power spectral density of a given instance of such process would be:

$$\Theta_{E_{NLI}}(f) = \sum_i |\mu_i|^2 \delta(f - if_0) \quad (61)$$

The overall RP average PSD is then simply:

$$G_{E_{NLI}}(f) = \mathbf{E}\{\Theta_{E_{NLI}}(f)\} = \sum_i \mathbf{E}\{|\mu_i|^2\} \delta(f - if_0) \quad (62)$$

where $\mathbf{E}$ is the statistical expectation operator. In other words, the average NLI power carried by a spectral line at frequency $if_0$ is found as $\mathbf{E}\{|\mu_i|^2\}$. Since the NLI power is the quantity of interest for performance evaluation, we now concentrate on the assessment of $\mathbf{E}\{|\mu_i|^2\}$.

*D. The NLI Power*

To evaluate $\mathbf{E}\{|\mu_i|^2\}$ we need to focus on each specific frequency component $if_0$. We can then write:

$$\mathbf{E}\{|\mu_i|^2\} = \mathbf{E}\left\{-j\gamma f_0^{\frac{3}{2}} e^{-j\beta(if_0)z} e^{-j2\gamma P_{TX}z_{eff}(z)} e^{-\alpha z} \right.$$
$$\sum_{m,n,k \in \tilde{A}_i} \sqrt{G_{TX}(mf_0)G_{TX}(nf_0)G_{TX}(kf_0)} \xi_m \xi_n^* \xi_k$$
$$\frac{1-e^{-2\alpha z}e^{j\{\beta([m-n+k]f_0)-\beta(mf_0)+\beta(nf_0)-\beta(kf_0)\}z}}{2\alpha - j\{\beta([m-n+k]f_0)-\beta(mf_0)+\beta(nf_0)-\beta(kf_0)\}}$$
$$j\gamma f_0^{\frac{3}{2}} e^{j\beta(if_0)z} e^{j2\gamma P_{TX}z_{eff}(z)} e^{-\alpha z} \sum_{m',n',k' \in \tilde{A}_i} \sqrt{G_{TX}(m'f_0)G_{TX}(n'f_0)G_{TX}(k'f_0)}$$
$$\left. \xi_{m'} \xi_{n'}^* \xi_{k'} \frac{1-e^{-2\alpha z}e^{-j\{\beta([m'-n'+k']f_0)-\beta(m'f_0)+\beta(n'f_0)-\beta(k'f_0)\}z}}{2\alpha + j\{\beta([m'-n'+k']f_0)-\beta(m'f_0)+\beta(n'f_0)-\beta(k'f_0)\}} \right\}$$
$$= \gamma^2 f_0^3 e^{-2\alpha z} \sum_{m,n,k \in \tilde{A}_i} \sum_{m',n',k' \in \tilde{A}_i} \mathbf{E}\{\xi_m \xi_n^* \xi_k \xi_{m'}^* \xi_{n'} \xi_{k'}^*\}$$
$$\sqrt{G_{TX}(mf_0)G_{TX}(nf_0)G_{TX}(kf_0)} \sqrt{G_{TX}(m'f_0)G_{TX}(n'f_0)G_{TX}(k'f_0)}$$
$$\frac{1-e^{-2\alpha z}e^{j\{\beta([m-n+k]f_0)-\beta(mf_0)+\beta(nf_0)-\beta(kf_0)\}z}}{2\alpha - j\{\beta([m-n+k]f_0)-\beta(mf_0)+\beta(nf_0)-\beta(kf_0)\}}$$
$$\frac{1-e^{-2\alpha z}e^{-j\{\beta([m'-n'+k']f_0)-\beta(m'f_0)+\beta(n'f_0)-\beta(k'f_0)\}z}}{2\alpha + j\{\beta([m'-n'+k']f_0)-\beta(m'f_0)+\beta(n'f_0)-\beta(k'f_0)\}}$$
(63)

From the last member of the above formula, we immediately see that the factor $e^{-j2\gamma P_{Tx}z_{eff}(z)}$ that arises from the action of the source term component which is proportional to the total transmitted power, given by Eq. (43), is completely irrelevant. This is because it is a pure phase-shift term, independent of the index $i$, and therefore it cancels out when the absolute value squared of $\mu_i$ is taken in Eq. (63). For this reason henceforth we will implicitly neglect this contribution, also in the context of the dual-polarization calculations outlined in Section IV-E.

The double summation within Eq. (63) gives rise to various classes of products. Most of such products, however, average to zero. Specifically, every time in the average

$$\mathbf{E}\{\xi_m \xi_n^* \xi_k \xi_{m'}^* \xi_{n'} \xi_{k'}^*\} \qquad (64)$$

one index appears only once, then the overall average goes to zero. This is because, as discussed, all the $\xi_n$'s in Eq. (13) are zero-mean and independent of one another. A thorough discussion of the various classes of terms arising from (64) is supplied in Appendix A. It turns out that the only classes of terms that are significant for large values of $W$ are those whereby:

$$m = m', \quad n = n', \quad k = k', \quad m \neq n, n \neq k, m \neq k$$
$$m = k', \quad n = n', \quad k = m', \quad m \neq n, n \neq k, m \neq k$$
(65)

for which it is:
$$\mathbf{E}\{\xi_m \xi_n^* \xi_k \xi_{m'}^* \xi_{n'} \xi_{k'}^*\} = \mathbf{E}\{|\xi_m|^2\} \mathbf{E}\{|\xi_n|^2\} \mathbf{E}\{|\xi_k|^2\} = 1.$$
(66)

We report here an approximation of Eq. (63), after all averaging has been carried out, derived in Appendix A, which is asymptotically exact for $W \to \infty$ (that is, since $f_0 = R_s/W$, for $f_0 \to 0$ in Eq. (17)):

$$\mathbf{E}\{|\mu_i|^2\} \approx 2\gamma^2 f_0^3 e^{-2\alpha z}$$
$$\sum_m \sum_k G_{Tx}(mf_0) G_{Tx}(kf_0) G_{Tx}([m+k-i]f_0)$$
$$\left| \frac{1-e^{-2\alpha z}e^{j\{\beta([m+k-i]f_0)-\beta(mf_0)+\beta(if_0)-\beta(kf_0)\}z}}{2\alpha - j\{\beta([m+k-i]f_0)-\beta(mf_0)+\beta(if_0)-\beta(kf_0)\}} \right|^2$$
(67)

Note that Eq. (67) is virtually coincident with Eq. (63) already for $W > 1000$, given the parameters of typical optical links. Note also that, remarkably, after averaging, the double summation over the set of triples $A_i$ simplifies to a double summation over two *independent* scalar indices.

By substituting Eq. (67) into Eq. (62) we obtain the PSD characterization of NLI noise, for a single polarization and a single span:

$$G_{E_{NLI}}(f) = 2\gamma^2 f_0^3 e^{-2\alpha z} \sum_i \delta(f - if_0)$$
$$\sum_m \sum_k G_{Tx}(mf_0) G_{Tx}(kf_0) G_{Tx}([m+k-i]f_0)$$
$$\left| \frac{1-e^{-2\alpha z}e^{j\{\beta([m+k-i]f_0)-\beta(mf_0)+\beta(if_0)-\beta(kf_0)\}z}}{2\alpha - j\{\beta([m+k-i]f_0)-\beta(mf_0)+\beta(if_0)-\beta(kf_0)\}} \right|^2$$
(68)

In the next two sections we upgrade this expression to dual-polarization and then to multiple spans.

### E. Accounting for Dual Polarization

To account for dual-polarization, we first need to suitably rewrite the Tx signal model. In this paper we assume PM transmission, with independent modulation on the two polarizations. We also assume that the Tx PSD is the same over either polarization. Such formats as PM-QPSK, PM-16QAM or PM-64QAM comply with these assumptions. Then, the transmitted signal is simply the juxtaposition of two single-polarization signals:

$$\vec{E}(0,f) = E_x(0,f)\hat{x} + E_y(0,f)\hat{y}$$

$$E_x(0,f) = \sqrt{f_0 \frac{G_{Tx}(f)}{2}} \sum_{n=-\infty}^{+\infty} \xi_{x,n} \delta(f - nf_0) \quad (69)$$

$$E_y(0,f) = \sqrt{f_0 \frac{G_{Tx}(f)}{2}} \sum_{n=-\infty}^{+\infty} \xi_{y,n} \delta(f - nf_0)$$

where the overhanging arrow indicates that the electric field is now a vector quantity. Note the factor $1/2$ multiplying $G_{Tx}(f)$. The reason for its presence is that this way we can still write the overall transmitted power through Eq. (21), simply by defining:

$$G_{Tx}(f) = G_{Tx,x}(f) + G_{Tx,y}(f) \quad (70)$$

where $G_{Tx,x}(f)$ and $G_{Tx,y}(f)$ are the PSDs of the signal on the $x$ and $y$ polarizations, respectively.

We then use the Manakov equation to model dual-polarization non-linear propagation. In frequency-domain we have:

$$\frac{\partial}{\partial z} E_x(z,f) = -j\beta(f)E_x(z,f) - \alpha E_x(z,f) +$$
$$-j\gamma \frac{8}{9}\left[ E_x(z,f) * E_x^*(z,-f) + E_y(z,f) * E_y^*(z,-f) \right] * E_x(z,f)$$

$$\frac{\partial}{\partial z} E_y(z,f) = -j\beta(f)E_y(z,f) - \alpha E_y(z,f) +$$
$$-j\gamma \frac{8}{9}\left[ E_x(z,f) * E_x^*(z,-f) + E_y(z,f) * E_y^*(z,-f) \right] * E_y(z,f)$$

(71)

Note that the two polarizations are coupled through the Kerr term. As it was done for the NLSE in Eq. (28) and (29), we assume the Kerr term to be a source term, calculated assuming linear propagation:

$$\frac{\partial}{\partial z} E_x(z,f) = -j\beta(f)E_x(z,f) - \alpha E_x(z,f) + Q_{NLI_x}(z,f)$$

$$\frac{\partial}{\partial z} E_y(z,f) = -j\beta(f)E_y(z,f) - \alpha E_y(z,f) + Q_{NLI_y}(z,f)$$

(72)

where:

$$Q_{NLI_x}(z,f) = -j\gamma \frac{8}{9} E_{LIN,x}(z,f) * E_{LIN,x}^*(z,-f) * E_{LIN,x}(z,f) +$$
$$-j\gamma \frac{8}{9} E_{LIN,y}(z,f) * E_{LIN,y}^*(z,-f) * E_{LIN,x}(z,f)$$

$$Q_{NLI_y}(z,f) = -j\gamma \frac{8}{9} E_{LIN,y}(z,f) * E_{LIN,y}^*(z,-f) * E_{LIN,y}(z,f) +$$
$$-j\gamma \frac{8}{9} E_{LIN,x}(z,f) * E_{LIN,x}^*(z,-f) * E_{LIN,y}(z,f)$$

(73)

The linear solutions for the field polarization components are completely independent and are of course similar to Eq. (50):

$$E_{LIN,x}(z,f) = E_x(0,f)e^{-j\beta(f)z - \alpha \cdot z}$$
$$E_{LIN,y}(z,f) = E_y(0,f)e^{-j\beta(f)z - \alpha \cdot z} \quad (74)$$

Here we have neglected the frequency-flat phase-shift due to the source term components proportional to the total transmission power, similar to that shown in Eq. (43), which we had included in the linear field solution for the single-polarization case Eqs. (47) and (50). We do so because, as commented in the previous subsection, they give rise to a frequency-independent phase-shifts which turn out to be completely irrelevant to the final result, that is to the PSD of NLI. Instead, their presence complicates the notation substantially.

Therefore, using Eq. (74) and the definitions in Eq. (73), it is possible to exploit the same procedure outlined in Eqs. (52)-(57) to derive the source terms. Their final expressions are very similar to Eq. (57). Specifically, for $Q_{NLI_x}(z,f)$ we have:

$$Q_{NLI_x}(z,f) = -j\gamma 2^{-\frac{3}{2}} \frac{8}{9} f_0^{\frac{3}{2}} e^{-3\alpha z}$$
$$\sum_{m=-\infty}^{\infty} \sum_{n=-\infty}^{\infty} \sum_{k=-\infty}^{\infty} \sqrt{G_{Tx}(mf_0)G_{Tx}(nf_0)G_{Tx}(kf_0)} \cdot$$
$$\left( \xi_{x,m}\xi_{x,n}^*\xi_{x,k} + \xi_{y,m}\xi_{y,n}^*\xi_{x,k} \right) \cdot$$
$$e^{-j[\beta(mf_0) - \beta(nf_0) + \beta(kf_0)]z} \delta\left(f - [m-n+k]f_0\right)$$

(75)

The differences between Eq. (75) and Eq. (57) are: the presence in the former of a factor $2^{-3/2}$ arising from the cube of the factor $2^{-1/2}$ inserted into Eq. (69); the factor $8/9$ that multiplies the Kerr term in the Manakov equation; two products of three RVs rather than just one product, which represent same-polarization and cross-polarization beatings. The expression for the other source term $Q_{NLI_y}(z,f)$ is immediately found by swapping the subscripts $x, y$ in Eq. (75).

We can then compute the NLI fields $E_{NLI_x}(0,f)$ and $E_{NLI_y}(0,f)$ following the same procedure as in Eqs. (58)-(59). The result for $E_{NLI_x}(0,f)$ is:

$$E_{NLI_x}(z,f) = \sum_{i=-\infty}^{\infty} \delta(f-if_0) \cdot \left[ -j\gamma 2^{-\frac{3}{2}} \frac{8}{9} f_0^{\frac{3}{2}} e^{-\alpha z} \right.$$

$$e^{-j\beta(if_0)z} \sum_{m,n,k\in\tilde{A}_i} \sqrt{G_{Tx}(mf_0)G_{Tx}(nf_0)G_{Tx}(kf_0)}$$

$$\left(\xi_{x,m}\xi_{x,n}^*\xi_{x,k} + \xi_{y,m}\xi_{y,n}^*\xi_{x,k}\right)$$

$$\left. \frac{1-e^{-2\alpha z}e^{j\{\beta([m-n+k]f_0)-\beta(mf_0)+\beta(nf_0)-\beta(kf_0)\}z}}{2\alpha - j\{\beta([m-n+k]f_0)-\beta(mf_0)+\beta(nf_0)-\beta(kf_0)\}} \right]$$

(76)

The result for $E_{NLI_y}(0,f)$ can be found again by swapping the subscripts $x$, $y$ in Eq. (32).

Note that $E_{NLI_x}(0,f)$ and $E_{NLI_y}(0,f)$ can be written as:

$$E_{NLI_x}(z,f) = \sum_i \mu_{x,i} \delta(f-if_0)$$

$$E_{NLI_y}(z,f) = \sum_i \mu_{y,i} \delta(f-if_0) \qquad (77)$$

similar to Eq. (60). We now want to find the average PSD of the dual-polarization NLI field:

$$\vec{E}_{NLI}(z,f) = E_{NLI_x}(z,f)\hat{x} + E_{NLI_y}(z,f)\hat{y} =$$
$$= \sum_i \left(\mu_{x,i}\hat{x} + \mu_{y,i}\hat{y}\right)\delta(f-if_0) \qquad (78)$$

Such PSD has a form similar to Eq. (62):

$$G_{E_{NLI}}(f) = \sum_i \left[ \mathbf{E}\{|\mu_{x,i}|^2\} + \mathbf{E}\{|\mu_{y,i}|^2\} \right] \delta(f-if_0)$$

(79)

where, similar to Eq. (63):

$$E\{|\mu_{x,i}|^2\} = \mathbf{E}\left\{ -j\gamma 2^{-\frac{3}{2}} \frac{8}{9} f_0^{\frac{3}{2}} e^{-j\beta(if_0)z} e^{-\alpha z} \right.$$

$$\sum_{m,n,k\in\tilde{A}_i} \sqrt{G_{Tx}(mf_0)G_{Tx}(nf_0)G_{Tx}(kf_0)} \left(\xi_{x,m}\xi_{x,n}^*\xi_{x,k} + \xi_{y,m}\xi_{y,n}^*\xi_{x,k}\right)$$

$$\frac{1-e^{-2\alpha z}e^{j\{\beta([m-n+k]f_0)-\beta(mf_0)+\beta(nf_0)-\beta(kf_0)\}z}}{2\alpha - j\{\beta([m-n+k]f_0)-\beta(mf_0)+\beta(nf_0)-\beta(kf_0)\}}$$

$$j\gamma 2^{-\frac{3}{2}} \frac{8}{9} f_0^{\frac{3}{2}} e^{j\beta(if_0)z} e^{-\alpha z} \sum_{m',n',k'\in\tilde{A}_i} \sqrt{G_{Tx}(m'f_0)G_{Tx}(n'f_0)G_{Tx}(k'f_0)}$$

$$\left(\xi_{x,m'}^*\xi_{x,n'}\xi_{x,k'}^* + \xi_{y,m'}^*\xi_{y,n'}\xi_{x,k'}^*\right)$$

$$\left. \frac{1-e^{-2\alpha z}e^{-j\{\beta([m'-n'+k']f_0)-\beta(m'f_0)+\beta(n'f_0)-\beta(k'f_0)\}z}}{2\alpha + j\{\beta([m'-n'+k']f_0)-\beta(m'f_0)+\beta(n'f_0)-\beta(k'f_0)\}} \right\} =$$

$$= \frac{8}{81}\gamma^2 f_0^3 e^{-2\alpha z} \sum_{m,n,k\in\tilde{A}_i} \sum_{m',n',k'\in\tilde{A}_i}$$

$$\left[ \mathbf{E}\{\xi_{x,m}\xi_{x,n}^*\xi_{x,k}\xi_{x,m'}^*\xi_{x,n'}\xi_{x,k'}^*\} + \mathbf{E}\{\xi_{x,m}\xi_{x,n}^*\xi_{x,k}\xi_{y,m'}^*\xi_{y,n'}\xi_{x,k'}^*\} \right.$$

$$\left. + \mathbf{E}\{\xi_{y,m}\xi_{y,n}^*\xi_{x,k}\xi_{x,m'}^*\xi_{x,n'}\xi_{x,k'}^*\} + \mathbf{E}\{\xi_{y,m}\xi_{y,n}^*\xi_{x,k}\xi_{y,m'}^*\xi_{y,n'}\xi_{x,k'}^*\} \right]$$

$$\sqrt{G_{Tx}(mf_0)G_{Tx}(nf_0)G_{Tx}(kf_0)}\sqrt{G_{Tx}(m'f_0)G_{Tx}(n'f_0)G_{Tx}(k'f_0)}$$

$$\frac{1-e^{-2\alpha z}e^{j\{\beta([m-n+k]f_0)-\beta(mf_0)+\beta(nf_0)-\beta(kf_0)\}z}}{2\alpha - j\{\beta([m-n+k]f_0)-\beta(mf_0)+\beta(nf_0)-\beta(kf_0)\}}$$

$$\frac{1-e^{-2\alpha z}e^{-j\{\beta([m'-n'+k']f_0)-\beta(m'f_0)+\beta(n'f_0)-\beta(k'f_0)\}z}}{2\alpha + j\{\beta([m'-n'+k']f_0)-\beta(m'f_0)+\beta(n'f_0)-\beta(k'f_0)\}}$$

(80)

The main difference with respect to Eq. (63), apart from the leading factor $8/81$, is the rather complex sum of statistical averages. A thorough discussion of such averaging process is reported in Appendix A. Here we summarize its results.

The first average in Eq. (80), that is:

$$\mathbf{E}\{\xi_{x,m}\xi_{x,n}^*\xi_{x,k}\xi_{x,m'}^*\xi_{x,n'}\xi_{x,k'}^*\} \qquad (81)$$

contains all $x$-polarization RVs and therefore it behaves exactly like the single average in Eq. (63), i.e., according to Eqs. (65)-(66). It accounts for same-polarization NLI.

The second and third averages in Eq. (80) are always zero or are irrelevant.

The fourth average, that is:

$$\mathbf{E}\{\xi_{y,m}\xi_{y,n}^*\xi_{x,k}\xi_{y,m'}^*\xi_{y,n'}\xi_{x,k'}^*\} \qquad (82)$$

is the one that accounts for cross-polarization NLI. Its contribution is, however, smaller than the contribution of the average of Eq. (81). This is because, looking at the two index value conditions listed in Eq. (65), we have that for:

$$m=m', \quad n=n', \quad k=k', \quad m\neq n, n\neq k, m\neq k \qquad (83)$$

both Eq. (81) and Eq. (82) evaluate to 1, whereas when:

$$m=k', \quad n=n', \quad k=m', \quad m\neq n, n\neq k, m\neq k \qquad (84)$$

Eq. (81) is still 1 but Eq. (82) evaluates to zero. In all other index combinations, both averages are zero.

After all averaging has been carried out, it is possible to derive an approximation to Eq. (80) which, similarly to Eq. (67), is asymptotically exact for $W \to \infty$ (see Appendix A):

$$\mathbf{E}\left\{\left|\mu_{x,i}\right|^2\right\} \approx \frac{8}{27}\gamma^2 f_0^3 e^{-2\alpha z} \cdot$$

$$\cdot \sum_m \sum_k G_{Tx}(mf_0) G_{Tx}(kf_0) G_{Tx}([m+k-i]f_0)$$

$$\left|\frac{1-e^{-2\alpha z}e^{j\{\beta([m+k-i]f_0)-\beta(mf_0)+\beta(if_0)-\beta(kf_0)\}z}}{2\alpha - j\{\beta([m+k-i]f_0)-\beta(mf_0)+\beta(if_0)-\beta(kf_0)\}}\right|^2 \tag{85}$$

Regarding the $y$-polarization term in Eq. (79), that is:

$$\mathbf{E}\left\{\left|\mu_{y,i}\right|^2\right\},$$

it is identical to Eq. (80), simply with the $x, y$ subscripts swapped. Its asymptotic approximation is identical to Eq. (85). Putting the two results together and recalling Eq. (79), we can finally write:

$$G_{\tilde{E}_{NLI}}(f) = \frac{16}{27}\gamma^2 f_0^3 e^{-2\alpha z} \sum_i \delta(f - if_0) \cdot$$
$$\sum_m \sum_k G_{TX}(mf_0) G_{TX}(kf_0) G_{TX}([m+k-i]f_0)$$
$$\left|\frac{1-e^{-2\alpha z}e^{j\{\beta([m+k-i]f_0)-\beta(mf_0)+\beta(if_0)-\beta(kf_0)\}z}}{2\alpha - j\{\beta([m+k-i]f_0)-\beta(mf_0)+\beta(if_0)-\beta(kf_0)\}}\right|^2 \tag{86}$$

This equation provides an analytical closed-form expression for the PSD of two-polarization NLI noise after one span of fiber. Note that, due to the behavior of the averages in Eq. (80) and in its homologue expressing $E\left\{\left|\mu_{y,i}\right|^2\right\}$, discussed above, it turns out that the contribution to NLI power of *same-polarization* beat terms is responsible for $2/3$ of the total NLI power, whereas the contribution of *cross-polarization* beat terms accounts for the remaining $1/3$. This is the case, independently of any link or system parameter. Also, the two contributions, i.e., same- and cross-polarization NLI, have exactly the same PSD.

### F. The Transition to a Frequency-Continuous Spectrum

As mentioned in [1], Appendix D-A, the discrete-summation based equations, such as (86), can be turned into integral equations, by replacing the sums with integrals. Greater details about the procedure will be reported in a forthcoming version of this document. In essence, once a formula like Eq. (68) or (86) are found, then it is possible to let $f_0 \to 0$, which is equivalent to making the delta-like spectrum of the signal increasingly thicker with spectral lines. Formally, what we compute is, for the dual-polarization case:

$$\lim_{f_0 \to 0}\{\text{Eq. (86)}\} \tag{87}$$

which yields:

$$G_{\tilde{E}_{NLI}}(f) = \frac{16}{27}\gamma^2 e^{-2\alpha z} \cdot$$
$$\int_{-\infty}^{\infty}\int_{-\infty}^{\infty} G_{Tx}(f_1) G_{Tx}(f_2) G_{Tx}(f_1+f_2-f) \cdot$$
$$\left|\frac{1-e^{-2\alpha z}e^{j\{\beta(f_1+f_2-f)-\beta(f_1)+\beta(f)-\beta(f_2)\}z}}{2\alpha - j\{\beta(f_1+f_2-f)-\beta(f_1)+\beta(f)-\beta(f_2)\}}\right|^2 \mathrm{d}f_1 \mathrm{d}f_2 \tag{88}$$

Note that the limit in (87) should in fact be viewed in the sense of distributions, that is: "the integral of Eq. (86) over any finite frequency interval in $f$ tends to be equal to the integral of Eq. (88) over any finite frequency interval in $f$, for $f_0 \to 0$".

Eq. (88) is more elegant than Eq. (86) and, above all, it lends itself to attempts at solving the integral analytically.

### G. Accounting for Multiple Identical Spans

The procedure with which multiple spans are tackled is outlined in [1], Appendix D-A. More similar details about the calculations can be found in Appendix B, where we derive the NLI field for the multiple different spans.

If identical spans of homogenous fibers are assumed, with lumped amplifiers exactly compensating for the loss of each span (including the last span), then the resulting NLI field at the end of the link $E_{NLI}(N_s L_s, f)$ is given by:

$$E_{NLI}(N_s L_s, f) =$$
$$= E_{NLI}^{(1)}(N_s L_s, f) + E_{NLI}^{(2)}(N_s L_s, f) +$$
$$+ E_{NLI}^{(3)}(N_s L_s, f) + \ldots + E_{NLI}^{(N_s)}(N_s L_s, f)$$
$$= -j\gamma f_0^{\frac{3}{2}} e^{-j2\gamma P_{Tx} N_s L_{s,eff}} \sum_{m,n,k \in \tilde{A}_i} e^{-j\beta([m-n+k]f_0)N_s L_s} \cdot$$
$$\sqrt{G_{Tx}(mf_0) G_{Tx}(nf_0) G_{Tx}(kf_0)} \cdot$$
$$\left[1 + e^{j\left[\beta([m-n+k]f_0)-\beta(mf_0)+\beta(nf_0)-\beta(kf_0)\right] \cdot L_s} +\right.$$
$$+ e^{j\left[\beta([m-n+k]f_0)-\beta(mf_0)+\beta(nf_0)-\beta(kf_0)\right] \cdot 2L_s} + \ldots$$
$$\left.\ldots + e^{j\left[\beta([m-n+k]f_0)-\beta(mf_0)+\beta(nf_0)-\beta(kf_0)\right] \cdot (N_s-1)L_s}\right] \cdot$$
$$\frac{1-e^{-2\alpha L_s} e^{j\{\beta([m-n+k]f_0)-\beta(mf_0)+\beta(nf_0)-\beta(kf_0)\}L_s}}{2\alpha - j\{\beta([m-n+k]f_0)-\beta(mf_0)+\beta(nf_0)-\beta(kf_0)\}} \cdot$$
$$\xi_m \xi_n^* \xi_k \delta\left(f - [m-n+k]f_0\right) \tag{89}$$

where $L_s$ is the span length, $N_s$ is the number of spans and:

$$L_{s,eff} = \frac{1-\exp(-2\alpha L_s)}{2\alpha} \tag{90}$$

is the effective length related to the physical length $L_s$. Also, $E_{NLI}^{(h)}(N_s L_s, f)$ is the NLI field produced in the $h$-th span, propagated to the end of the link. Eq. (89) simply adds up the NLI contributions due to each span, after propagating them to the end of the link, including a last amplifier that makes up for the loss of the last span. It is interesting to see that all the contributions $E_{NLI}^{(h)}(N_s L_s, f)$ are formally identical, except for a phase factor, which gives rise to the factor:

$$\begin{bmatrix} 1 + e^{j[\beta([m-n+k]f_0)-\beta(mf_0)+\beta(nf_0)-\beta(kf_0)]\cdot L_s} + \\ + e^{j[\beta([m-n+k]f_0)-\beta(mf_0)+\beta(nf_0)-\beta(kf_0)]\cdot 2L_s} + \ldots \\ \ldots + e^{j[\beta([m-n+k]f_0)-\beta(mf_0)+\beta(nf_0)-\beta(kf_0)]\cdot (N_s-1)L_s} \end{bmatrix} \tag{91}$$

This factor represents the coherent interference of the NLI contributions, due to each span, at the end of the link. Eq. (91) is a truncated geometric series and can be summed up analytically:

$$\begin{bmatrix} 1 + e^{j\Phi} + e^{j2\Phi} + \ldots + e^{j(N_s-1)\Phi} \end{bmatrix} = \sum_{h=0}^{N_s-1} \left(e^{j\Phi}\right)^h =$$
$$= \frac{1-e^{jN_s\Phi}}{1-e^{j\Phi}} = e^{j(N_s-1)\Phi/2} \frac{\sin(N_s\Phi/2)}{\sin(\Phi/2)} \tag{92}$$

Inserting Eq. (92) into Eq. (89), one gets:

$$E_{NLI}(N_s L_s, f) =$$
$$-j\gamma f_0^{\frac{3}{2}} e^{-j2\gamma P_{Tx} N_s L_{s,eff}} \sum_{m,n,k\in\tilde{A}_i} e^{-j\beta([m-n+k]f_0)N_s L_s}$$
$$\sqrt{G_{Tx}(mf_0)G_{Tx}(nf_0)G_{Tx}(kf_0)}$$
$$e^{j[\beta([m-n+k]f_0)-\beta(mf_0)+\beta(nf_0)-\beta(kf_0)]\cdot(N_s-1)L_s/2}$$
$$\frac{\sin\left(\left[\beta([m-n+k]f_0)-\beta(mf_0)+\beta(nf_0)-\beta(kf_0)\right]\cdot N_s L_s/2\right)}{\sin\left(\left[\beta([m-n+k]f_0)-\beta(mf_0)+\beta(nf_0)-\beta(kf_0)\right]\cdot L_s/2\right)}$$
$$\frac{1-e^{-2\alpha L_s}e^{j\{\beta([m-n+k]f_0)-\beta(mf_0)+\beta(nf_0)-\beta(kf_0)\}L_s}}{2\alpha-j\{\beta([m-n+k]f_0)-\beta(mf_0)+\beta(nf_0)-\beta(kf_0)\}}$$
$$\xi_m \xi_n^* \xi_k \delta(f-[m-n+k]f_0) \tag{93}$$

This result corresponds to Eq. (46) in Appendix D-A of [1]. The $\sin(N_s\Phi/2)/\sin(\Phi/2)$ is sometimes called "phased array" factor because it formally looks like a phased-array antenna radiation diagram. Such interference effect, with similar analytical form, was first pointed out in the context of conventional FWM calculations [9], [10].

For a detailed discussion of the implications of the phased array factor on NLI noise accumulation see [4], Section XI-C.

From Eq. (93) onward, calculations follow as from Eq. (59) onward. The extension to dual polarization also follows the same procedure as shown from Eq. (69) onward.

The final result for the NLI PSD after $N_s$ identical spans, with dual-polarization, is remarkably similar to that of the single span, provided by Eq. (86). In fact, the only apparent difference is the presence of the phased-array term, squared:

$$G_{\bar{E}_{NLI}}(f) = \frac{16}{27}\gamma^2 f_0^3 \sum_i \delta(f-if_0)$$
$$\sum_m \sum_k G_{Tx}(mf_0)G_{Tx}(kf_0)G_{Tx}([m+k-i]f_0)$$
$$\left|\frac{1-e^{-2\alpha L_s}e^{j\{\beta([m+k-i]f_0)-\beta(mf_0)+\beta(if_0)-\beta(kf_0)\}L_s}}{2\alpha-j\{\beta([m+k-i]f_0)-\beta(mf_0)+\beta(if_0)-\beta(kf_0)\}}\right|^2$$
$$\frac{\sin^2\left(\left[\beta([m+k-i]f_0)-\beta(mf_0)+\beta(if_0)-\beta(kf_0)\right]\cdot N_s L_s/2\right)}{\sin^2\left(\left[\beta([m+k-i]f_0)-\beta(mf_0)+\beta(if_0)-\beta(kf_0)\right]\cdot L_s/2\right)} \tag{95}$$

Another minor difference is the disappearance of the factor $e^{-2\alpha z}$, due to the fact we have assumed optical amplifiers compensating for fiber loss all along, including at the end of the last span. This equation already represents a significant and practically usable result provided by the model.

From Eq. (95), we see again see that the factor $e^{-j2\gamma P_{Tx} N_s L_{s,eff}}$ present in Eq. (93) disappears completely and has no effect.

Such factor arises from the action of the source term component which is proportional to the total transmitted power, given by Eq. (43). However, since it is a pure phase-shift term, independent of the index $i$, it cancels out when the absolute value squared of Eq. (93) is taken. In the next section, this phase shift will be disregarded, for notational simplicity.

Similar to what was done in Section IV-F, a transition to a "continuous spectrum" is possible. Following the same procedure, the final result is:

$$G_{\bar{E}_{NLI}}(f) = \frac{16}{27}\gamma^2 \int_{-\infty}^{\infty}\int_{-\infty}^{\infty} G_{TX}(f_1)G_{TX}(f_2)G_{TX}(f_1+f_2-f)$$
$$\left|\frac{1-e^{-2\alpha L_s}e^{j\{\beta(f_1+f_2-f)-\beta(f_1)+\beta(f)-\beta(f_2)\}L_s}}{2\alpha-j\{\beta(f_1+f_2-f)-\beta(f_1)+\beta(f)-\beta(f_2)\}}\right|^2$$
$$\frac{\sin^2\left(\left[\beta(f_1+f_2-f)-\beta(f_1)+\beta(f)-\beta(f_2)\right]\cdot N_s L_s/2\right)}{\sin^2\left(\left[\beta(f_1+f_2-f)-\beta(f_1)+\beta(f)-\beta(f_2)\right]\cdot L_s/2\right)} df_1 df_2 \tag{96}$$

If we only consider the lower-order dispersion terms $\beta_2$ and $\beta_3$, the propagation constant $\beta(f)$ can be written as,

$$\beta(f) = \frac{\beta_2}{2}(2\pi f)^2 + \frac{\beta_3}{6}(2\pi f)^3$$
$$= 2\pi^2 \beta_2 f^2 + \frac{4}{3}\pi^3 \beta_3 f^3 \quad (G.1)$$

Then the term $\beta(f_1 + f_2 - f) - \beta(f_1) + \beta(f) - \beta(f_2)$ in Eq. (96) can be explicitly expressed as,

$$\beta(f_1 + f_2 - f) - \beta(f_1) + \beta(f) - \beta(f_2)$$
$$= \left(2\pi^2 \beta_2 (f_1 + f_2 - f)^2 + \frac{4}{3}\pi^3 \beta_3 (f_1 + f_2 - f)^3\right) +$$
$$-\left(2\pi^2 \beta_2 f_1^2 + \frac{4}{3}\pi^3 \beta_3 f_1^3\right) + \left(2\pi^2 \beta_2 f^2 + \frac{4}{3}\pi^3 \beta_3 f^3\right) +$$
$$-\left(2\pi^2 \beta_2 f_2^2 + \frac{4}{3}\pi^3 \beta_3 f_2^3\right)$$
$$= 2\pi^2 \beta_2 \left((f_1 + f_2 - f)^2 - f_1^2 + f^2 - f_2^2\right) +$$
$$+ \frac{4}{3}\pi^3 \beta_3 \left((f_1 + f_2 - f)^3 - f_1^3 + f^3 - f_2^3\right)$$
$$= 2\pi^2 \beta_2 \left(2(f_1 - f)(f_2 - f)\right)$$
$$+ \frac{4}{3}\pi^3 \beta_3 \left(3(f_1 + f_2)(f_1 - f)(f_2 - f)\right)$$
$$= 4\pi^2 (f_1 - f)(f_2 - f)\left[\beta_2 + \pi \beta_3 (f_1 + f_2)\right] \quad (G.2)$$

Substituting Eq. (G.2) into Eq. (96), we can obtain,

$$G_{\tilde{E}_{NLI}}(f) = \frac{16}{27}\gamma^2 \int_{-\infty}^{\infty}\int_{-\infty}^{\infty} G_{TX}(f_1) G_{TX}(f_2) G_{TX}(f_1 + f_2 - f)$$
$$\left|\frac{1 - e^{-2\alpha L_s} e^{j4\pi^2 (f_1 - f)(f_2 - f)[\beta_2 + \pi\beta_3(f_1 + f_2)]L_s}}{2\alpha - j4\pi^2 (f_1 - f)(f_2 - f)[\beta_2 + \pi\beta_3(f_1 + f_2)]}\right|^2 \quad (G.3)$$
$$\frac{\sin^2\left(2N_s \pi^2 (f_1 - f)(f_2 - f)[\beta_2 + \pi\beta_3(f_1 + f_2)]L_s\right)}{\sin^2\left(2\pi^2 (f_1 - f)(f_2 - f)[\beta_2 + \pi\beta_3(f_1 + f_2)]L_s\right)} df_1 df_2$$

Obviously, by setting $\beta_3 = 0$, (G.3) can be expressed as a function of $\beta_2$:

$$G_{\tilde{E}_{NLI}}(f) = \frac{16}{27}\gamma^2 \int_{-\infty}^{\infty}\int_{-\infty}^{\infty} G_{TX}(f_1) G_{TX}(f_2) G_{TX}(f_1 + f_2 - f)$$
$$\left|\frac{1 - e^{-2\alpha L_s} e^{j4\pi^2 (f_1 - f)(f_2 - f)\beta_2 L_s}}{2\alpha - j4\pi^2 (f_1 - f)(f_2 - f)\beta_2}\right|^2 \quad (G.4)$$
$$\frac{\sin^2\left(2N_s \pi^2 (f_1 - f)(f_2 - f)\beta_2 L_s\right)}{\sin^2\left(2\pi^2 (f_1 - f)(f_2 - f)\beta_2 L_s\right)} df_1 df_2$$

which is the so-called GNRF, or GN-model reference formula, as reported in [4]. In the following two sections IV-H and IV-I we will consider both $\beta_2$ and $\beta_3$.

### H. Accounting for Multiple Different Spans

Here we neglect the frequency-flat phase-shift due to the source term components proportional to the total transmission power, similar to that shown in Eq. (43), which we had included in the linear field solution for the single-polarization case, i.e., Eqs. (47) and (50). We do so because, as commented in the previous subsection, it gives rise to a frequency-independent phase-shift which turns out to be completely irrelevant to the final result, that is to the PSD of NLI.

In the following, we introduce the expression of the NLI field at the end of a link, assuming that the spans can have:
- different fiber lengths and types
- different amplification gain at the end of each span

We also assume that the signals that were present at launch propagate till the end of the link, without any channels being spilled out or inserted into the link. The details of the derivation are outlined in Appendix B.

The resulting NLI field is then given by:

$$E_{NLI}(L_{tot}, f) = \sum_{n_s=1}^{N_s} E_{NLI}^{(n_s)}(L_{tot}, f) =$$
$$= -jf_0^{\frac{3}{2}} \sum_i \delta(f - if_0) \sum_{m,n,k \in \tilde{A}_i} \xi_m \xi_n^* \xi_k$$
$$\sqrt{G_{Tx}(mf_0) G_{Tx}(nf_0) G_{Tx}(kf_0)}$$
$$e^{-j2\pi^2 f_0^2 i^2 \sum_{n_s=1}^{N_s}\left(\beta_{2,n_s} L_{s,n_s} + \frac{2}{3}\pi i f_0 \beta_{3,n_s} L_{s,n_s} + \beta_{DCU,n_s}\right)}$$
$$\sum_{n_s=1}^{N_s} \gamma_{n_s} \cdot \prod_{n_s'=1}^{n_s-1} g_{n_s'}^{3/2} e^{-3\alpha_{n_s'} L_{s,n_s'}} \cdot \prod_{n_s'=n_s}^{N_s} g_{n_s'}^{1/2} e^{-\alpha_{n_s'} L_{s,n_s'}}$$
$$e^{j4\pi^2 f_0^2 (k-n)(m-n) \sum_{n_s'=1}^{n_s-1}\left(\beta_{2,n_s'} L_{s,n_s'} + \pi(m+k) f_0 \beta_{3,n_s'} L_{s,n_s'} + \beta_{DCU,n_s'}\right)}$$
$$\frac{1 - e^{-2\alpha_{n_s} L_{s,n_s}} e^{j4\pi^2 f_0^2 (k-n)(m-n)\left[\beta_{2,n_s} + \pi(m+k) f_0 \beta_{3,n_s}\right] L_{s,n_s}}}{2\alpha_{n_s} - j4\pi^2 f_0^2 (k-n)(m-n)\left[\beta_{2,n_s} + \pi(m+k) f_0 \beta_{3,n_s}\right]}$$

$$(97)$$

where:

$$L_{tot} = \sum_{n_s=1}^{N_s} L_{s,n_s} \qquad (98)$$

and where a number of quantities appear, which refer to the $n_s$-th span: $L_{s,n_s}$ its length; $\alpha_{n_s}$ its loss parameter; $\beta_{2,n_s}$ and $\beta_{3,n_s}$ its dispersion parameters; $\gamma_{n_s}$ its non-linearity coefficient; $\beta_{DCU,n_s}$ a possible amount of lumped dispersion in $\left[\text{ps}^2\right]$ placed at the end of the span; $g_{n_s}$ the power gain of the EDFA placed at the end of the $n_s$-th span. Then, following again the squaring and averaging procedure outlined in Section IV-D and IV-E, we get:

$$G_{\vec{E}_{NLI}}(f) = \frac{16}{27} f_0^3 \sum_i \delta(f - if_0)$$

$$\sum_m \sum_k G_{Tx}(mf_0) G_{Tx}(kf_0) G_{Tx}([m+k-i]f_0)$$

$$\left| \sum_{n_s=1}^{N_s} \gamma_{n_s} \cdot \prod_{n_s'=1}^{n_s-1} g_{n_s'}^{3/2} e^{-3\alpha_{n_s'} L_{s,n_s'}} \cdot \prod_{n_s'=n_s}^{N_s} g_{n_s'}^{1/2} e^{-\alpha_{n_s'} L_{s,n_s'}} \right.$$

$$e^{j4\pi^2 f_0^2 (m-i)(k-i) \sum_{n_s'=1}^{n_s-1} \left(\beta_{2,n_s'} L_{s,n_s'} + \pi(m+k)f_0 \beta_{3,n_s'} L_{s,n_s'} + \beta_{DCU,n_s'}\right)}$$

$$\left. \frac{1 - e^{-2\alpha_{n_s} L_{s,n_s}} e^{j4\pi^2 f_0^2 (m-i)(k-i)\left[\beta_{2,n_s} + \pi(m+k)f_0 \beta_{3,n_s}\right] L_{s,n_s}}}{2\alpha_{n_s} - j4\pi^2 f_0^2 (m-i)(k-i)\left[\beta_{2,n_s} + \pi(m+k)f_0 \beta_{3,n_s}\right]} \right|^2$$

(99)

Carrying out the transition to continuous frequency, as it was done in Section IV-F, we get:

$$G_{\vec{E}_{NLI}}(f) = \frac{16}{27} \int_{-\infty}^{\infty} \int_{-\infty}^{\infty} G_{Tx}(f_1) G_{Tx}(f_2) G_{Tx}(f_1 + f_2 - f)$$

$$\left| \sum_{n_s=1}^{N_s} \gamma_{n_s} \cdot \prod_{n_s'=1}^{n_s-1} g_{n_s'}^{3/2} e^{-3\alpha_{n_s'} L_{s,n_s'}} \cdot \prod_{n_s'=n_s}^{N_s} g_{n_s'}^{1/2} e^{-\alpha_{n_s'} L_{s,n_s'}} \right.$$

$$e^{j4\pi^2 (f_1-f)(f_2-f) \sum_{n_s'=1}^{n_s-1} \left(\beta_{2,n_s'} L_{s,n_s'} + \pi(f_1+f_2)\beta_{3,n_s'} L_{s,n_s'} + \beta_{DCU,n_s'}\right)}$$

$$\left. \frac{1 - e^{-2\alpha_{n_s} L_{s,n_s}} e^{j4\pi^2 (f_1-f)(f_2-f)\left[\beta_{2,n_s} + \pi\beta_{3,n_s}(f_1+f_2)\right] L_{s,n_s}}}{2\alpha_{n_s} - j4\pi^2 (f_1-f)(f_2-f)\left[\beta_{2,n_s} + \pi\beta_{3,n_s}(f_1+f_2)\right]} \right|^2 df_2 df_1$$

(100)

Note that, if one assumes that span loss is exactly compensated for at the end of each span, then $g_{n_s} e^{-2\alpha_{n_s} L_{s,n_s}} = 1$, $\forall n_s$, and therefore the gain-loss related products appearing in Eq. (100) simplify:

$$\prod_{n_s'=1}^{n_s-1} g_{n_s'}^{3/2} e^{-3\alpha_{n_s'} L_{s,n_s'}} = 1$$

$$\prod_{n_s'=n_s}^{N_s} g_{n_s'}^{1/2} e^{-\alpha_{n_s'} L_{s,n_s'}} = 1$$

making the overall equation substantially simpler.

It should be noted that, if DCU compensation is close to full compensation, such as in dispersion-managed systems, some of the assumptions the model is based on would not hold anymore. Therefore, the results might be unreliable. Research is ongoing on this aspect, so please check for publications by the same authors of this document for possible developments.

If the incoherent NLI accumulation assumption is made, as an approximation to Eq. (100), the following result is found:

$$G_{\vec{E}_{NLI}}(f) = \sum_{n_s=1}^{N_s} \gamma_{n_s}^2 \cdot \prod_{n_s'=1}^{n_s-1} g_{n_s'}^3 e^{-6\alpha_{n_s'} L_{s,n_s'}} \cdot \prod_{n_s'=n_s}^{N_s} g_{n_s'} e^{-2\alpha_{n_s'} L_{s,n_s'}} \cdot$$

$$\frac{16}{27} \int_{-\infty}^{\infty} \int_{-\infty}^{\infty} G_{Tx}(f_1) G_{Tx}(f_2) G_{Tx}(f_1 + f_2 - f) \cdot$$

$$\left| \frac{1 - e^{-2\alpha_{n_s} L_{s,n_s}} e^{j4\pi^2 (f_1-f)(f_2-f)\left[\beta_{2,n_s} + \pi\beta_{3,n_s}(f_1+f_2)\right] L_{s,n_s}}}{2\alpha_{n_s} - j4\pi^2 (f_1-f)(f_2-f)\left[\beta_{2,n_s} + \pi\beta_{3,n_s}(f_1+f_2)\right]} \right|^2 df_2 df_1$$

(101)

So far we have assumed that the channels launched at the input are not spilled out or replaced by others. In future flexible networks it may be that some channels are removed or added at any span, while others are allowed to propagate undisturbed. In this scenario there is no real "start" or "end" of the link. The "first" and "last" span can be arbitrarily tagged so, for any specific channel.

Unfortunately, writing down a general formula for this case is difficult, because different parts of the signal spectrum would have to carry along different phase and amplitude "histories", which would combine non-linearly in a complex way. Investigation is ongoing on the possibility of achieving a reasonable-complexity general formula for this case.

### I. The Distributed Amplification Case

Another interesting generalization can be operated versus the span loss/gain profile. If we employ the same procedure as in section IV-H, and also consider the distributed gain, we can get the NLI PSD at the end of the link,

$$G_{\tilde{E}_{NLI}}(f) = \frac{16}{27} \int_{-\infty}^{\infty} \int_{-\infty}^{\infty} G_{Tx}(f_1) G_{Tx}(f_2) G_{Tx}(f_1+f_2-f)$$

$$\left| \sum_{n_s=1}^{N_s} \gamma_{n_s} \cdot \prod_{n'_s=1}^{n_s-1} g_{n'_s}^{3/2} e^{-3\alpha_{n'_s} L_{s,n'_s}} e^{\int_0^{L_{s,n'_s}} 3\hat{g}_{n'_s}(\zeta)d\zeta} \prod_{n'_s=n_s}^{N_s} g_{n'_s}^{1/2} e^{-\alpha_{n'_s} L_{s,n'_s}} e^{\int_0^{L_{s,n'_s}} \hat{g}_{n'_s}(\zeta)d\zeta} \right.$$

$$e^{j4\pi^2(f_1-f)(f_2-f)\sum_{n'_s=1}^{n_s-1}\left(\beta_{2,n'_s} L_{s,n'_s} + \pi(f_1+f_2)\beta_{3,n'_s} L_{s,n'_s} + \beta_{DCU,n'_s}\right)}$$

$$\left. \int_0^{L_{s,n_s}} e^{\int_0^z 2\hat{g}_{n_s}(\zeta)d\zeta} e^{-2\alpha_{n_s} z} e^{j4\pi^2(f_1-f)(f_2-f)\left(\beta_{2,n_s}+\pi\beta_{3,n_s}(f_1+f_2)\right)z} dz \right|^2 df_2 df_1$$

(I.1)

where $\hat{g}$ is the (field) local gain coefficient [1/km] due to distributed amplification. Note that gain $\hat{g}$ can be a function of the longitudinal spatial coordinate.

Next we assume all identical spans, with span loss exactly compensated for by amplification, and we do not consider the lumped dispersion any more. If so, the NLI PSD at the end of the link can be re-written as:

$$G_{\tilde{E}_{NLI}}(f) = \frac{16}{27} \gamma^2 L_{eff}^2 \int_{-\infty}^{\infty} \int_{-\infty}^{\infty} \rho(f_1, f_2, f) \cdot$$

$$\frac{\sin^2\left(2N_s \pi^2 (f_1-f)(f_2-f)[\beta_2+\pi\beta_3(f_1+f_2)]L_s\right)}{\sin^2\left(2\pi^2 (f_1-f)(f_2-f)[\beta_2+\pi\beta_3(f_1+f_2)]L_s\right)} \cdot$$

$$G_{Tx}(f_1) G_{Tx}(f_2) G_{Tx}(f_1+f_2-f) df_1 df_2$$

(102)

where:

$$\rho(f_1, f_2, f) =$$

$$= \left| \int_0^{L_s} e^{\int_0^z 2\hat{g}(\zeta)d\zeta} e^{-2\alpha z} e^{j4\pi^2(f_1-f)(f_2-f)[\beta_2+\pi\beta_3(f_1+f_2)]z} dz \right|^2 \Big/ L_{eff}^2$$

(103)

$$L_{eff} = \int_0^{L_s} e^{\int_0^z 2\hat{g}(\zeta)d\zeta} e^{-2\alpha z} dz \qquad (104)$$

The factor $\rho$ defined by (103) is in fact the non-degenerate FWM efficiency, normalized so that its maximum value is 1, of the beating of three spectral lines, positioned at frequencies $f_1$, $f_2$, and $f_3 = f_1+f_2-f$, which create a beat line at frequency $f$, interacting over a span. Note that the maximum of $\rho$ is found at $\rho(0,0,0)$.

The definitions (102)-(104) can be used to numerically assess NLI for any fiber gain/loss profile. They can also be used a starting point to derive analytical results for certain specific distributed-amplification scenarios. In particular, expressions (103)-(104) can be given approximate closed-form solutions in two significant cases.

One is that of ideal distributed amplification, whereby distributed gain equals loss at all points along the fiber. This case was discussed at length in [4], where a full derivation of $\rho$ was shown, together with the subsequent calculation of an approximate solution for the corresponding NLI PSD. Although very difficult to achieve in practice, this scenario is of interest because it represents the best performing amplification setup and therefore it can be used to establish fundamental capacity limits. Such limits were studied in depth in [13].

Another interesting case, this one of direct practical importance, is that of backward-pumped Raman amplification.

*1) Backward-pumped Raman amplification*

We assume that a single counter-propagating pump is used or that a multi-pump set-up is approximated through an equivalent single-pump. We neglect pump depletion. We consider the resulting Raman gain as frequency independent. Then, the local Raman power gain generated by such set-up at a location $z$ along the fiber span is given by the standard formula:

$$2\hat{g}(z) = C_R P_{p0} e^{2\alpha_p z} \qquad (105)$$

where $C_R$ is the Raman gain coefficient [1/(W km)], $\alpha_p$ [1/km] is the (field) attenuation coefficient of the fiber at the Raman pump frequency and $P_{p0}$ is the power of the pump at $z=0$, i.e., at the start of the span. Note that, since the pump is injected at the end of the span and propagates backwards, its power-profile actually grows exponentially in the forward $z$ direction, as shown by (105).

We then solve the inner integral in (103):

$$e^{\int_0^z 2\hat{g}(\zeta)d\zeta} e^{-2\alpha z} = e^{-2\alpha z} e^{C_R P_{p0} \int_0^z e^{2\alpha_p \zeta} d\zeta}$$

$$= e^{-2\alpha z} e^{C_R P_{p0} \left[e^{2\alpha_p z} - 1\right]/2\alpha_p}$$

(106)

Introducing (106) into (103), we then get:

$$\rho(f_1, f_2, f) =$$

$$= \left| e^{-\frac{C_R P_{p0}}{2\alpha_p}} \int_0^{L_s} e^{-2\alpha z} e^{\frac{C_R P_{p0}}{2\alpha_p} e^{2\alpha_p z}} e^{j4\pi^2(f_1-f)(f_2-f)[\beta_2+\pi\beta_3(f_1+f_2)]z} dz \right|^2 \Big/ L_{eff}^2$$

(107)

We then remark that the integral inside the absolute value sign has the form:

$$\int_0^z e^{ae^{bx}} e^{-cx} dx \qquad (108)$$

with:

$$\text{Re}\{a\} > 0 \quad \text{Im}\{a\} = 0$$
$$\text{Re}\{b\} > 0 \quad \text{Im}\{b\} = 0$$
$$\text{Re}\{c\} > 0 \quad \text{Im}\{c\} \neq 0$$

We then carry out a change of integration variable:

$$u = e^{bx} \quad x = \frac{\ln u}{b} \quad dx = \frac{1}{bu} du \quad (109)$$
$$\text{Re}\{u\} > 0 \quad \text{Im}\{u\} = 0$$

so that we obtain:

$$\int_0^z e^{ae^{bx}} e^{-cx} dx = \int_1^{e^{bz}} e^{au} e^{-c\frac{\ln u}{b}} \frac{1}{bu} du$$
$$= \frac{1}{b}\int_1^{e^{bz}} u^{-1} e^{au} e^{-c\frac{\ln u}{b}} du = \frac{1}{b}\int_1^{e^{bz}} u^{-1} e^{au} e^{\ln u^{-c/b}} du \quad (110)$$
$$= \frac{1}{b}\int_1^{e^{bz}} u^{-1} u^{-c/b} e^{au} du = \frac{1}{b}\int_1^{e^{bz}} u^{-c/b-1} e^{au} du$$

With the aid of the following further substitution:

$$r = \frac{c}{b}$$

we can finally write:

$$\int_0^{L_s} e^{-2\alpha z} e^{\frac{C_R P_{p_0}}{2\alpha_p} e^{2\alpha_p z}} e^{j4\pi^2 (f_1-f)(f_2-f)[\beta_2+\pi\beta_3(f_1+f_2)]z} dz =$$
$$= \frac{1}{b}\int_1^{e^{bL_s}} u^{-(1+r)} e^{au} du \quad (111)$$

where:

$$a = \frac{C_R P_{p_0}}{2\alpha_p}; \quad b = 2\alpha_p;$$
$$r = \frac{\alpha}{\alpha_p} - \frac{j2\pi^2(f_1-f)(f_2-f)[\beta_2+\pi\beta_3(f_1+f_2)]}{\alpha_p}$$
$$\text{Re}\{a\} > 0 \quad \text{Im}\{a\} = 0$$
$$\text{Re}\{b\} > 0 \quad \text{Im}\{b\} = 0$$
$$\text{Re}\{r\} > 0 \quad \text{Im}\{r\} \neq 0$$

The right-hand side integral in (111) has the analytical solution:

$$\frac{1}{b}\int_1^{e^{bL_s}} u^{-(1+r)} e^{au} du = \frac{1}{b}\left[-u^{-r} \cdot (-au)^r \Gamma(-r,-au)\right]_1^{e^{bL_s}}$$
$$= \frac{1}{b}\left[(-a)^r \left\{\Gamma(-r,-a) - \Gamma(-r,-ae^{bL_s})\right\}\right]$$
(113)

where $\Gamma(x_1,x_2)$ is the upper incomplete Gamma function, defined as:

$$\Gamma(x,y) = \int_y^\infty w^{x-1} e^{-w} dw \quad (114)$$

Taking result (113) into account, from (111) and (112) we then have:

$$\int_0^{L_s} e^{-2\alpha z} e^{\frac{C_R P_{p_0}}{2\alpha_p} e^{2\alpha_p z}} e^{j4\pi^2(f_1-f)(f_2-f)[\beta_2+\pi\beta_3(f_1+f_2)]z} dz =$$
$$= \frac{1}{2\alpha_p}\left\{\left(-\frac{C_R P_{p_0}}{2\alpha_p}\right)^{\left[\frac{\alpha}{\alpha_p} - \frac{j2\pi^2(f_1-f)(f_2-f)[\beta_2+\pi\beta_3(f_1+f_2)]}{\alpha_p}\right]} \cdot \right.$$
$$\left[\Gamma\left(-\left[\frac{\alpha}{\alpha_p} - \frac{j2\pi^2(f_1-f)(f_2-f)[\beta_2+\pi\beta_3(f_1+f_2)]}{\alpha_p}\right], -\frac{C_R P_{p_0}}{2\alpha_p}\right)\right.$$
$$\left.\left.-\Gamma\left(-\left[\frac{\alpha}{\alpha_p} - \frac{j2\pi^2(f_1-f)(f_2-f)[\beta_2+\pi\beta_3(f_1+f_2)]}{\alpha_p}\right], -\frac{C_R P_{p_0}}{2\alpha_p} e^{2\alpha_p L_s}\right)\right]\right\}$$
(115)

Finally, substituting (115) into (103), we get:

$$\rho(f_1,f_2,f) =$$
$$= \frac{1}{L_{eff}^2}\left|\frac{e^{-C_R P_{p_0}/2\alpha_p}}{2\alpha_p}\left(-\frac{C_R P_{p_0}}{2\alpha_p}\right)^{\left[\frac{\alpha}{\alpha_p} - \frac{j2\pi^2(f_1-f)(f_2-f)[\beta_2+\pi\beta_3(f_1+f_2)]}{\alpha_p}\right]} \cdot \right.$$
$$\left[\Gamma\left(-\left[\frac{\alpha}{\alpha_p} - \frac{j2\pi^2(f_1-f)(f_2-f)[\beta_2+\pi\beta_3(f_1+f_2)]}{\alpha_p}\right], -\frac{C_R P_{p_0}}{2\alpha_p}\right)\right.$$
$$\left.\left.-\Gamma\left(-\left[\frac{\alpha}{\alpha_p} - \frac{j2\pi^2(f_1-f)(f_2-f)[\beta_2+\pi\beta_3(f_1+f_2)]}{\alpha_p}\right], -\frac{C_R P_{p_0}}{2\alpha_p} e^{2\alpha_p L_s}\right)\right]\right|^2$$
(116)

Regarding the effective length $L_{eff}$, we point out that its derivation is very similar to that of $\rho$, provided that one sets $f = f_1 = f_2 = 0$. Then using formula (115), one immediately finds:

$$L_{eff} = e^{-C_R P_{p_0}/2\alpha_p} \frac{1}{2\alpha_p} \cdot$$

$$\left\{ \left(-\frac{C_R P_{p_0}}{2\alpha_p}\right)^{\frac{\alpha}{\alpha_p}} \left[ \Gamma\left(-\frac{\alpha}{\alpha_p}, -\frac{C_R P_{p_0}}{2\alpha_p}\right) - \Gamma\left(-\frac{\alpha}{\alpha_p}, -\frac{C_R P_{p_0}}{2\alpha_p} e^{2\alpha_p L_s}\right) \right] \right\}$$

(117)

Unfortunately, so far it has not been possible to proceed further in the analytical integration of GNRF with backward-pumped Raman amplification. Therefore, (116) and (117) must be inserted back into (102) and numerical integration must be carried out. Note that using hyperbolic coordinates, as shown in [4], may substantially ease the task.

Note that Eqs. (116) and (117) were already reported in [4], but no derivation was given there for lack of space.

## V. APPROXIMATE ANALYTICAL SOLUTION FOR THE CASE OF NON-IDENTICAL CHANNELS

In this version of the document, we reproduce only the final results for this case. A subsequent version will incorporate the analytical derivation.

We start out by defining the system scenario. We assume that the channels in the comb:
- may have different power
- may have different bandwidth
- may have uneven spacing

We assume for now identical spans, with span loss exactly compensated for by span amplification. This assumption will be removed later on.

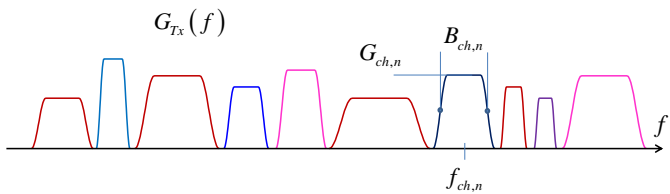

**Figure V-1 Example of a possible WDM overall transmission spectrum $G_{WDM}(f)$. Each channel can have a different -3dB bandwidth $B_{ch}$, a different center frequency $f_{ch}$ and a different flat-top power spectral density $G_{ch}$. The different colors are only meant to stress channel diversity.**

An example of the WDM spectrum of one such transmission system is provided in Figure V-1, under the assumption that channels have a raised-cosine power spectral density with relatively small roll-off. This assumption is not critical for the derivation of the approximate NLI PSD, but the accuracy of the approximation progressively degrades as the roll-off increases, or if the spectrum is substantially different than flat-top (such as a sinc-like spectrum).

As shown in Figure V-1, we take as channel bandwidth the -3dB (half-height) bandwidth, $B_{ch,n}$. The index $n$ runs from 1 to the number of channels $N_{ch}$. If root-raised cosine ISI-free pulses are used to generate the channels, then $B_{ch,n}$ coincides with the symbol rate $R_{s,n}$. The center frequency and the flat-top PSD value are represented by the quantities $f_{ch,n}$ and $G_{ch,n}$. We then write:

$$G_{Tx}(f) = \sum_{n=1}^{N_{ch}} G_{ch,n}(f) = \sum_{n=1}^{N_{ch}} G_{ch,n} \cdot g_{ch,n}(f)$$

(118)

where the quantity $g_{ch,n}(f)$ is a normalized PSD and $N_{ch}$ is the number of channels. Note that we impose:

$$\max\{g_{ch,n}(f)\} = 1 \qquad (119)$$

In previous sections we concentrated on the NLI spectrum appearing on top of the center channel. Here we generalize and write a formula which provides an approximation of the NLI PSD appearing at the center frequency of anyone channel, say, the $i$-th channel. It is given by the following expression:

$$G_{\bar{E}_{NLI}}(f_{ch,i}) \approx \frac{16}{27} \gamma^2 L_{eff}^2 \sum_{n=1}^{N_{ch}} G_{ch,n} G_{ch,n} G_{ch,i} \cdot (2 - \delta_{ni}) \psi_{n,i}$$

(120)

where the coefficients $\psi_{n,i}$ are given by:

$$\psi_{n,i} = \int_{-\infty}^{\infty} \int_{-\infty}^{\infty} \rho(f_1, f_2, f_{ch,i}) \cdot \chi(f_1, f_2, f_{ch,i}) \cdot$$
$$\cdot g_{ch,n}(f_1) g_{ch,i}(f_2) g_{ch,n}(f_1 + f_2 - f_{ch,i}) df_1 df_2$$

(121)

We have used the following shorthand for the phased-array term:

$$\chi(f_1, f_2, f) = \frac{\sin^2\left(2N_s \pi^2 (f_1 - f)(f_2 - f)\beta_2 L_s\right)}{\sin^2\left(2\pi^2 (f_1 - f)(f_2 - f)\beta_2 L_s\right)}$$

(122)

Note also that $\delta_{ni} = 1$ if $n = i$ and $\delta_{ni} = 0$ otherwise.

To obtain the approximation (120), only the SCI and XCI components of the overall NLI noise were taken into account. The MCI component was neglected. For an explanation of this terminology and a discussion of why, in general, MCI is negligible, see [4], Sect. VI.

It is interesting to remark that if MCI had been included, then (120) would have consisted of a triple summation. As a result,

an overall number of terms proportional to $N_{ch}^3$ would have had to be evaluated, each one similar to (121). By excluding MCI, the triple summation reduces to a single summation, with the overall approximation complexity becoming proportional to only $N_{ch}$.

Under the assumption of *incoherent accumulation* (see Sect. 0), Eq. (121) can shed the phased-array factor $\chi$. NLI noise then simply accumulates linearly vs. the number of spans $N_s$:

$$\psi_{n,i} = N_s \int_{-\infty}^{\infty} \int_{-\infty}^{\infty} \rho(f_1, f_2, f_{ch,i}) \cdot$$
$$\cdot g_{ch,n}(f_1) g_{ch,i}(f_2) g_{ch,n}(f_1 + f_2 - f_{ch,i}) df_1 df_2 \quad (123)$$

Further assuming to be dealing with lumped amplification, Eq. (123) can be shown to be well approximated by:

$$\psi_{n,i} \approx N_s \frac{\operatorname{asinh}\left(\pi^2[2\alpha]^{-1}|\beta_2|\left[f_{ch,n} - f_{ch,i} + B_{ch,n}/2\right]B_{ch,i}\right)}{4\pi(2\alpha)^{-1}|\beta_2|} -$$

$$-N_s \frac{\operatorname{asinh}\left(\pi^2[2\alpha]^{-1}|\beta_2|\left[f_{ch,n} - f_{ch,i} - B_{ch,n}/2\right]B_{ch,i}\right)}{4\pi(2\alpha)^{-1}|\beta_2|}, \quad n \neq i$$

$$\psi_{i,i} \approx N_s \frac{\operatorname{asinh}\left(\frac{\pi^2}{2}|\beta_2|[2\alpha]^{-1} B_{ch,i}^2\right)}{2\pi|\beta_2|[2\alpha]^{-1}}$$

(124)

where "asinh" is a hyperbolic arc-sine. Note that the contributions with $n \neq i$ are XCI, whereas the contribution for $n = i$ is SCI.

Regarding the accuracy of the incoherent accumulation assumption, we point out that, as argued in [4], *coherent* accumulation affects mostly SCI, that is, NLI noise created by the channel under test onto itself. As a result, an easy and rather accurate correction strategy consists of taking coherent accumulation into account only for $\psi_{i,i}$. If so, $\psi_{i,i}$ would change into:

$$\psi_{i,i} \approx N_s^{1+\varepsilon} \frac{\operatorname{asinh}\left(\frac{\pi^2}{2}|\beta_2|[2\alpha]^{-1} B_{ch,i}^2\right)}{2\pi|\beta_2|[2\alpha]^{-1}}$$

(125)

where:

$$\varepsilon \approx \frac{3}{10} \cdot \log_e\left(1 + \frac{6}{L_s} \frac{(2\alpha)^{-1}}{\operatorname{asinh}\left(\frac{1}{2}\pi^2|\beta_2|(2\alpha)^{-1} B_{ch,i}^2\right)}\right)$$

(126)

The above formula (126) for the exponent correction $\varepsilon$ showing up in Eq. (125) was found in [4]. In practice, the coherent SCI contribution (124)-(125) is identical to formulas (13) and (23) from [4].

As a further remark, if all channels are assumed to be equal bandwidth and power, and having equal spacing, then (120) and (124) become identical to Eq. (40) in [4], which indeed deals with WDM systems having identical equispaced channels.

### A. Non-Identical Spans

A final important sub-case is when spans are not identical. To be able to deal with such case in closed form, one must accept the incoherent accumulation assumption. Then the NLI noise spectrum at the end of the system can be simply calculated as the sum of the NLI noise spectra produced in each single span, taking however into account the loss and gain experienced in each span.

With this in mind, combining the results from Sect. 0, Eq. (101), with Eqs. (120) and (124), we can then obtain the rather general closed-form result:

$$G_{\bar{E}_{NLI}}(f_{ch,i}) = \frac{16}{27} \sum_{n_s=1}^{N_s} \gamma_{n_s}^2 L_{eff,n_s}^2 \cdot$$

$$\prod_{n'_s=1}^{n_s-1} g_{n'_s}^3 e^{-6\alpha_{n'_s} L_{s,n'_s}} \cdot \prod_{n'_s=n_s}^{N_s} g_{n'_s} e^{-2\alpha_{n'_s} L_{s,n'_s}} \cdot \quad (127)$$

$$\sum_{n=1}^{N_{ch}} G_{ch,n} G_{ch,n} G_{ch,i} \cdot (2 - \delta_{ni}) \psi_{n,i,n_s}$$

$$\psi_{n,i,n_s} \approx \frac{\operatorname{asinh}\left(\pi^2\left[2\alpha_{n_s}\right]^{-1}|\beta_{2,n_s}|\left[f_{ch,n} - f_{ch,i} + B_{ch,n}/2\right]B_{ch,i}\right)}{4\pi(2\alpha_{n_s})^{-1}|\beta_{2,n_s}|} -$$

$$-\frac{\operatorname{asinh}\left(\pi^2\left[2\alpha_{n_s}\right]^{-1}|\beta_{2,n_s}|\left[f_{ch,n} - f_{ch,i} - B_{ch,n}/2\right]B_{ch,i}\right)}{4\pi(2\alpha_{n_s})^{-1}|\beta_{2,n_s}|}, \quad n \neq i$$

(128)

$$\psi_{i,i,n_s} \approx \frac{\operatorname{asinh}\left(\frac{\pi^2}{2}|\beta_{2,n_s}|\left[2\alpha_{n_s}\right]^{-1} B_{ch,i}^2\right)}{2\pi|\beta_{2,n_s}|\left[2\alpha_{n_s}\right]^{-1}}$$

(129)

### VI. CONCLUSION

This document provides an in-depth treatment of the derivation of the GN-model of non-linear interference presented in [1]-[4]. In its present form, it is a dynamic document, whereby some parts will be expanded in the future to provide even more details, so that the readers can find it easier to re-derive and possibly extend or modify the model themselves. Original material is included as well.

It should also be mentioned that ample room is still available for further research on the GN model. For instance: closed-form analytical results which are more accurate and/or encompass a wider range of system scenarios, than presented in [4]; a broader exploration of the validity envelope of
the model, especially towards very low dispersion values and low symbol rates, where signal "gaussianization" after launch is slow; model extensions aiming at encompassing dispersion-managed systems and mixed-fiber systems, and so on. In all these respects, having a detailed derivation available, greatly eases the undertaking of future research, which is in fact the main goal of this paper.

### A. Appendix A

A key passage in the derivation of the NLI PSD is the statistical averaging that takes place in Eq. (63) for the single-polarization case, and in Eq. (80) for the dual-polarization case. In this version of this document, the discussion is limited to the single-polarization case. We will extend the discussion to the dual-polarization case in a future version.

The statistical features of Eq. (63) are concentrated within the following set of averages:

$$E\left\{\xi_m \xi_n^* \xi_k \xi_{m'}^* \xi_{n'} \xi_{k'}^*\right\}$$
$$\{(m,n,k): m-n+k = i\} \quad (B1)$$
$$\{(m',n',k'): m'-n'+k' = i\}$$

where $i$ is the index identifying the frequency $if_0$ at which the NLI PSD is to be evaluated. To ease the discussion, we will now make some assumptions which do not imply any relevant loss of generality.

We first assume that the number of channels in the comb is odd. This clearly identifies a *center channel* in the comb itself. We also define a *center frequency* for the grid, $f_{center}$ which we assume to coincide with the nominal carrier frequency of the center channel of the comb. We assume that the channels are uniformly spaced with spacing $f_{sp}$. Then, looking at Eq. (16), we identify a frequency range outside of which $E(f) = 0$, based in turn on the frequency range outside of which where the Tx spectrum $G_{Tx}(f) = 0$. We call such range:

$$B_{WDM} \equiv \left[f_{low}, f_{high}\right] \quad (B2)$$

and we impose that this frequency range is symmetric with respect to the center frequency, that is:

$$\left|f_{low} - f_{center}\right| = \left|f_{high} - f_{center}\right| \quad (B3)$$

We then define an equispaced frequency grid $f_i = i \cdot f_0$ across the whole range $B_{WDM}$. Mathematically, we can write it as:

$$f_i = f_{center} + i \cdot f_0 \quad (B4)$$

with $f_0 = f_{sp} / K$. The integer $K$ is the number of grid points falling between two channels (including one of the two carriers). In the following, for compactness, we will always drop $f_{center}$, so that the WDM comb center channel frequency will conventionally be zero.

In Eq. (B1) each of the six indices $m,n,k,m',n',k'$ identifies a signal frequency on the grid which, combining with others, creates an interference term at frequency $i \cdot f_0$. These indices therefore need only span the set of frequencies where the Tx signal is present, that is they need to cover $B_{WDM}$ but do not need to extend further. Hence:

$$m,n,k,m',n',k' \in \left\lceil \frac{f_{low}}{f_0} \right\rceil \ldots \left\lfloor \frac{f_{high}}{f_0} \right\rfloor \quad (B5)$$

The resulting total number of grid frequencies addressed by each index is then $M$, defined as:

$$M = \left\lfloor \frac{f_{high}}{f_0} \right\rfloor - \left\lceil \frac{f_{low}}{f_0} \right\rceil + 1 \quad (B6)$$

Notice that $M$ is always an odd number, because by construction $\left\lfloor f_{high}/f_0 \right\rfloor = -\left\lceil f_{low}/f_0 \right\rceil$. For convenience we also define a quantity:

$$\mathcal{M} = \left\lfloor M/2 \right\rfloor = \left\lfloor f_{high}/f_0 \right\rfloor = -\left\lceil f_{low}/f_0 \right\rceil \quad (B7)$$

so that we can compactly re-write the index range as:

$$m,n,k,m',n',k' \in -\mathcal{M} \ldots \mathcal{M} \quad (B8)$$

Given Eq. (B8) and the relations between these indices and the index $i$ in Eq. (B6), it is easy to see that:

$$i \in -3\mathcal{M} \ldots 3\mathcal{M} \quad (B9)$$

That is, similarly to FWM, the NLI terms extend over a bandwidth which is triple with respect to the signal bandwidth. We now concentrate on the NLI terms which fall at the center frequency of the center channel, that is at $i = 0$. In other words we address the averages within the quantity $\mathbf{E}\left\{\left|\mu_0\right|^2\right\}$ in Eq.

(22). Such averages are the ones in Eq. (B1) for $i=0$. We will later extend the discussion to any value of $i$. The triples spanned by $m,n,k$ are:

$$A_0 = \{(m,n,k): m-n+k=0;\ m,n,k \in -M...M\}$$
(B10)

An identical relation can be written for $m',n',k'$. The denomination $A_0$ for this set was introduced in Eq. (21).

We can plot the entire set $A_0$ over a 2D space with axes $m,n$ and use $k$ as a parameter. In Fig. (B1) we show an example assuming $M=11$, that is $\mathcal{M}=5$. This value of $M$ is low and is chosen only for convenience. This choice does not imply any loss of generality in the following discussion.

An identical plot can be drawn for $m',n',k'$. The total number of triples shown in Fig. (B1) is exactly $3\mathcal{M}^2 + 3\mathcal{M} + 1$.

For convenience, we first classify the triples in subsets. To avoid introducing new terminology, we call such subsets according to the classical terminology of Kerr non-linear terms, with the *caveat* that their meaning only relates to the formally identical index sets spanned and not to the effect on transmission, which in our context is drastically different:

- ND-FWM triples: all three indices different (Fig. B2); defining $\mathcal{M} = \lfloor M/4 \rfloor$, there are exactly $[3\mathcal{M}^2 - \mathcal{M} - 2\mathcal{M}]$ such triples
- D-FWM triples: $m=k$, excluding $m=n=k=0$ (Fig. B3);, there are exactly $2\mathcal{M}$ such triples
- XPM triples: $m=n$ or $n=k$, excluding $m=n=k=0$ (Fig. B4); there are exactly $4\mathcal{M}$ such triples
- SPM term: a single triple $m=n=k=0$

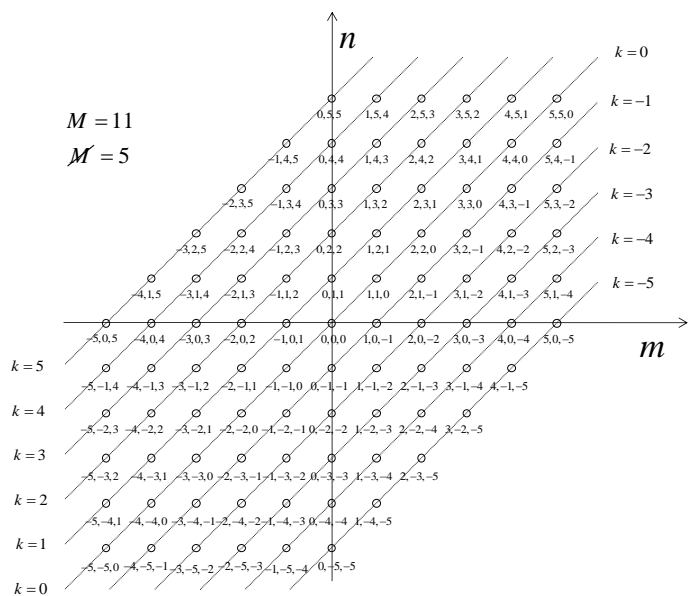

Fig. (B1): example of the graphical representation of the set of triples $m,n,k$ in $A_0$, from Eq. (B10), assuming $M=11$. Each dot is marked accordingly.

According to Eq. (B6), each average $E\{\xi_m \xi_n^* \xi_k \xi_{m'}^* \xi_{n'} \xi_{k'}^*\}$ involves a triple $(m,n,k)$ and a triple $(m',n',k')$. In other words, chosen a specific marker on the $A_0$ plane, it interacts with every marker on an identical plane. The resulting total number of averages of the type (B1) is therefore $(3\mathcal{M}^2 + 3\mathcal{M} + 1)^2$. Most of them, however, are zero, as we shall see below.

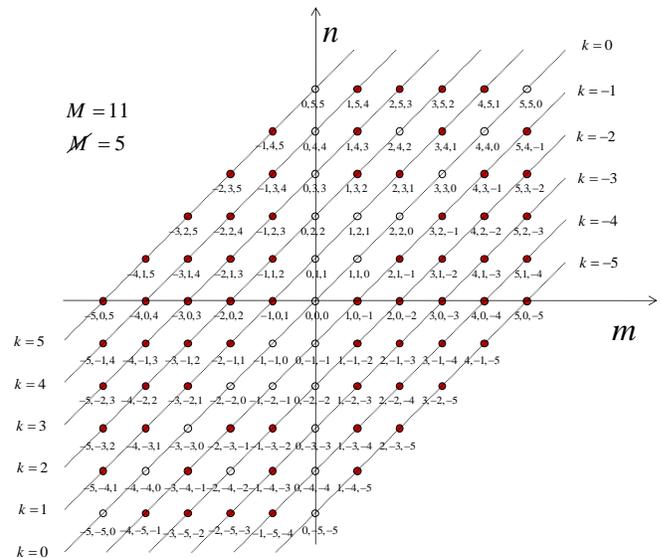

Fig. B2: Same as Fig. B1. Here the filled dots identify the ND-FWM triples, that is, those for which all three indices are different.

We start out by addressing a ND-FWM triple, combining with any other triple. We set off by looking at those triples for which $m \neq n \neq k$. For each one of these triples, there are only two possible configurations of the triple $(m',n',k')$ for which the average $E\{\xi_m \xi_n^* \xi_k \xi_{m'}^* \xi_{n'} \xi_{k'}^*\}$ is non-zero. They occur when:

$$m=m', n=n', k=k', \quad \text{or}$$
$$m=k', n=n', k=m' \quad \text{(B11)}$$

In both these instances, we have:

$$E\left\{\xi_m\xi_n^*\xi_k\xi_{m'}^*\xi_{n'}\xi_{k'}^*\right\} = E\left\{\left|\xi_m\right|^2\left|\xi_n\right|^2\left|\xi_k\right|^2\right\} =$$
$$= E\left\{\left|\xi_m\right|^2\right\}E\left\{\left|\xi_n\right|^2\right\}E\left\{\left|\xi_k\right|^2\right\} = 1 \quad \text{(B12)}$$

Otherwise, there is at least one index that occurs only once, so that at least one single-RV average factorizes, of the type: $E\{\xi_p\}$, which is zero and therefore causes the overall average to go to zero; or, all indices appear twice but the complex-conjugates do not match the non-complex conjugates, so that one or more average factorizes of the type $E\{\xi_p^2\}$, which also evaluates to zero, because by assumption the RV's $\xi_p$ are complex Gaussian RV's, that is they can be written as: $\xi_p = \xi_{R,p} + j\xi_{I,p}$, where the $\xi_{R,p}$, $\xi_{I,p}$, are independent zero-mean real Gaussian RV's. As a result:

$$E\{\xi_p^2\} = E\{(\xi_{R,p} + j\xi_{I,p})^2\} =$$
$$E\{\xi_{R,p}^2\} - E\{\xi_{I,p}^2\} + 2jE\{\xi_{R,p}\}E\{\xi_{I,p}\} = 0 \quad \text{(B13)}$$

Eventually all the ND-FWM triples $(m,n,k)$ in Fig. B2, interacting with all the similar triples $(m',n',k')$, generate exactly a total of $2\left[3M^2 - M - 2M\right]$ non-zero averages, whose value is actually 1.

We now look at D-FWM triples, as depicted in Fig. B3, for which $m = k$. All averages where one of such triples combines with any triple is zero, for the same reasons it so happened for the ND-FWM case. The only exception occurs again when conditions (B11) occur. This time however, the two conditions are degenerate because $m = k$, so that in fact only one non-zero average is found (and not two).

The result is again:

$$E\left\{\left|\xi_m\right|^2\right\}E\left\{\left|\xi_n\right|^2\right\}E\left\{\left|\xi_k\right|^2\right\} = 1 \quad \text{(B14)}$$

A total of exactly $2M$ averages whose value is 1 is generated by D-FWM triples. Note therefore that the D-FWM non-zero contributions grow only as $M$, whereas the number of non-zero ND-FWM contributions grow as $M^2$. Also, the strength of either contribution is limited. Therefore, as the number of spectral lines $M+1$ used to model the signal is increased, i.e. if $f_0$ is decreased, then clearly the D-FWM contribution becomes negligible vs. the ND-FWM contribution and can be neglected.

Regarding the XPM triples (Fig. B4), the detailed examination of all possible combinations of an XPM triple with all possible triples leads to concluding that the total number of non-zero contributions arising from XPM-related triples is $16M^2$. As a result, it appears that they could not be neglected, even by letting $M$ increase.

However, including XPM triples into the averaging process is not only cumbersome but also conceptually wrong. Each triple $(m,n,k)$ relates to one term of the summation carried by the source term of Eq. (33), shown again below for convenience:

$$Q_{NLI}(0,f) = -j\gamma f_0^{\frac{3}{2}}\sum_{i=-\infty}^{\infty}\delta(f - if_0)\cdot$$
$$\sum_{m,n,k\in A_i}\xi_m\xi_n^*\xi_k\sqrt{G_{Tx}(mf_0)G_{Tx}(nf_0)G_{Tx}(kf_0)} \quad \text{(B15)}$$

Although the signal model implies that the $\xi$'s are RVs, certain quantities related to them have non-random values. Specifically, as shown in Eqs. (37)-(39), the sum over the XPM terms leads to a non-statistical quantity, that is to the total signal power. In fact, restricting the sum over all XPM terms, we get:

$$Q_{NLI,XPM}(0,f) =$$
$$-j\gamma f_0^{\frac{1}{2}}\sum_{i=-\infty}^{\infty}\delta(f - if_0)\cdot\xi_i\sqrt{G_{Tx}(if_0)}\cdot \quad \text{(B16)}$$
$$2f_0\sum_{\substack{n=-M\\n\neq 0}}^{M}\left|\xi_n\right|^2 G_{Tx}(nf_0)$$

but clearly the last line is, according to Eq. (24)-(25):

$$2f_0\sum_{\substack{n=-M\\n\neq 0}}^{M}\left|\xi_n\right|^2 G_{Tx}(nf_0) \approx 2P_{Tx} \quad \text{(B17)}$$

The approximate equality stems from the fact that the SPM term $n = 0$ is excluded. However, clearly for increasing $M$ the above approximations converges to exact relationship, whereas, it can be shown that the contributions of the SPM triple are completely vanishing with respect to all other contributions. The above relationship was found at $z = 0$. However, it keeps on holding at any z, since as shown in Eq. (56) reproduced here for convenience:

$$Q_{NLI}(z,f) = -j\gamma f_0^{\frac{3}{2}} e^{-j2\beta_2\pi^2 f^2 z} e^{-3\alpha z} e^{-j2\gamma P_{Tx} z_{eff}(z)}$$

$$\sum_{i=-\infty}^{\infty} \delta(f - if_0) \sum_{m,n,k \in A_i} \sqrt{G_{Tx}(mf_0) G_{Tx}(nf_0) G_{Tx}(kf_0)}$$

$$\xi_m \xi_n^* \xi_k e^{j4\pi^2 \beta_2 f_0^2 (k-n)(m-n) z}$$

(B18)

the inner summation only adds a phase-shift factor, which makes it irrelevant as to its absolute-value squared. In addition, however, also notice that:

$$e^{j4\pi^2 \beta_2 f_0^2 (k-n)(m-n) z} = 1 \qquad (B19)$$

for $m = n$ or $k = n$. Therefore, relation (B17) keeps holding at any $z$.

To summarize, out of all the contributions to the averaging process, that of D-FWM is vanishing, as well as that of SPM, for increasing $M$. Regarding XPM, these contributions would not be vanishing but they can be dealt with separately, by converting them to a deterministic contribution. The only contribution that must be looked at is then that of ND-FWM, which was discussed before and for which all averages are 1.

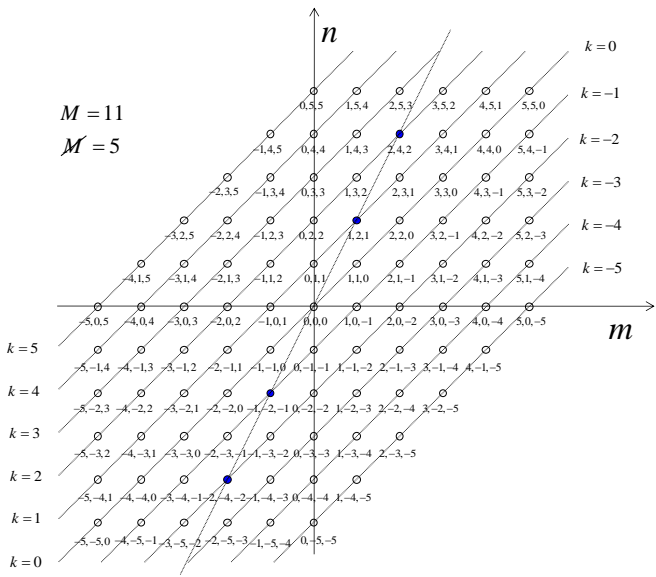

Fig. B3: Same as Fig. B1. Here the filled dots identify the D-FWM triples, that is those for which $m = k$.

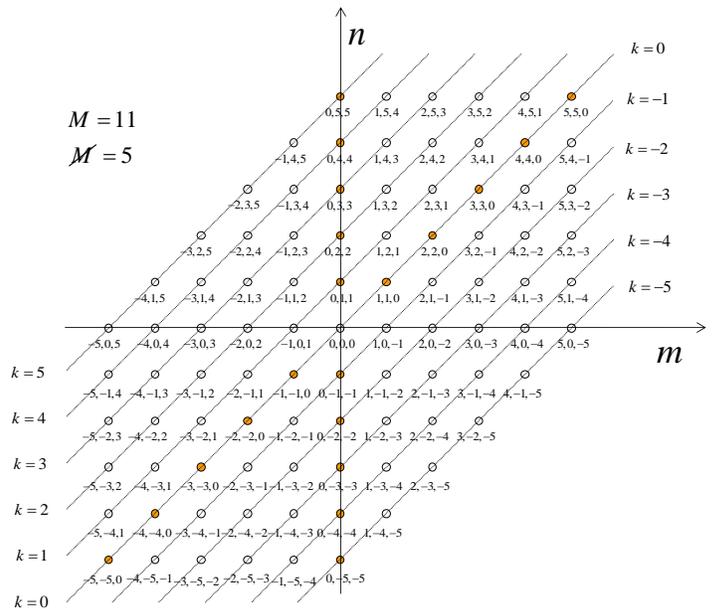

Fig. B4: Same as Fig. B1. Here the filled dots identify the XPM triples, that is those for which $m = n$ or $n = k$.

If this is taken into account, the averaging process yields the approximate expression Eq. (68). The approximation indeed stems from neglecting D-FWM and SPM and from the approximation in (B17), all of which however become negligible as $M$ is increased.

The above discussion was carried out referring mainly to the case $i = 0$, for simplicity. It can be shown that the conclusions reached for the case $i = 0$ hold true also for the case of a generic value of $i$.

Notice also that the transition to integral expressions is in fact based on assuming $M \to \infty$, so in this sense the integral expressions are no longer approximated expressions.

### B. Appendix B

For each span, say the $n_s$-th span, the signal propagates linearly in the first $n_s - 1$ spans, and generates NLI in the $n_s$-th span, then the NLI propagates linearly in the following spans. Here, we just consider the lower order fiber dispersion parameters $\beta_2$ and $\beta_3$.

For the $n_s$-th span, the input field is,

$$E_{LNI}^{(n_s)}\left(\sum_{n'_s=1}^{n_s-1} L_{s,n'_s}, f\right) =$$

$$= e^{-j2\pi^2 f^2 \sum_{n'_s=1}^{n_s-1}\left(\beta_{2,n'_s} L_{s,n'_s} + \frac{2}{3}\pi f \beta_{3,n'_s} L_{s,n'_s} + \beta_{DCU,n'_s}\right)} \cdot \text{(B20)}$$

$$\prod_{n'_s=1}^{n_s-1} g_{n'_s}^{1/2} e^{-\alpha_{n'_s} L_{s,n'_s}} \cdot E(0,f)$$

where the quantities referring to the $n_s$-th span are demonstrated in section IV-H, and the signal model $E(0,f)$ is shown in Eq. (16).

In the $n_s$-th span, the linear field is,

$$E_{LNI}^{(n_s)}(z,f) = e^{\Gamma(z,f)} \cdot E_{LNI}^{(n_s)}\left(\sum_{n'_s=1}^{n_s-1} L_{s,n'_s}, f\right)$$

$$= e^{-j2\pi^2 f^2 \left[\sum_{n'_s=1}^{n_s-1}\left(\beta_{2,n'_s} L_{s,n'_s} + \frac{2}{3}\pi f \beta_{3,n'_s} L_{s,n'_s} + \beta_{DCU,n'_s}\right) + \beta_{2,n_s} z + \frac{2}{3}\pi f \beta_{3,n_s} z\right]} \cdot$$

$$e^{-\alpha_{n_s} z} \prod_{n'_s=1}^{n_s-1} g_{n'_s}^{1/2} e^{-\alpha_{n'_s} L_{s,n'_s}} \cdot E(0,f)$$

(B21)

Then we can get the NLI noise term using Eq. (52),

$$Q_{NLI,\tilde{A}_i}^{(n_s)}(z,f) \approx$$

$$\approx -j\gamma_{n_s} E_{LNI}^{(n_s)}(z,f) * E_{LNI}^{(n_s)*}(z,-f) * E_{LNI}^{(n_s)}(z,f)$$

$$= -j\gamma_{n_s} f_0^{\frac{3}{2}} \prod_{n'_s=1}^{n_s-1} g_{n'_s}^{3/2} e^{-3\alpha_{n'_s} L_{s,n'_s}} \cdot e^{-3\alpha_{n_s} z} \cdot$$

$$\sum_{i=-\infty}^{\infty} \delta(f-if_0) \sum_{m,n,k \in \tilde{A}_i} \xi_m \xi_n^* \xi_k \cdot$$

$$e^{-j2\pi^2 f_0^2 \sum_{n'_s=1}^{n_s-1}\left(\left(\beta_{2,n'_s} L_{s,n'_s} + \beta_{DCU,n'_s}\right)(m^2-n^2+k^2) + \frac{2}{3}\pi \beta_{3,n'_s} L_{s,n'_s}(m^3-n^3+k^3)f_0\right)} \cdot$$

$$e^{-j2\pi^2 f_0^2 \left[\beta_{2,n_s} z(m^2-n^2+k^2) + \frac{2}{3}\pi \beta_{3,n_s} z(m^3-n^3+k^3)f_0\right]} \cdot$$

$$\sqrt{G_{TX}(mf_0) G_{TX}(nf_0) G_{TX}(kf_0)}$$

(B22)

Therefore, the nonlinear field in the $n_s$-th span is,

$$E_{NLI}^{(n_s)}(z,f) = e^{\Gamma(z,f)} \cdot \int_0^z e^{-\Gamma(z',f)} Q_{NLI,\tilde{A}}^{(n_s)}(z',f) dz'$$

$$= -j\gamma_{n_s} f_0^{\frac{3}{2}} \prod_{n'_s=1}^{n_s-1} g_{n'_s}^{3/2} e^{-3\alpha_{n'_s} L_{s,n'_s}} \cdot e^{-\alpha_{n_s} z}$$

$$\sum_{i=-\infty}^{\infty} \delta(f-if_0) e^{-j2\pi^2 i^2 f_0^2 \left(\beta_{2,n_s} + \frac{2}{3}\pi \beta_{3,n_s} if_0\right)z}$$

$$\sum_{m,n,k \in \tilde{A}_i} \xi_m \xi_n^* \xi_k \sqrt{G_{TX}(mf_0) G_{TX}(nf_0) G_{TX}(kf_0)}$$

$$e^{-j2\pi^2 f_0^2 \sum_{n'_s=1}^{n_s-1}\left(\left(\beta_{2,n'_s} L_{s,n'_s} + \beta_{DCU,n'_s}\right)(m^2-n^2+k^2) + \frac{2}{3}\pi \beta_{3,n'_s} L_{s,n'_s}(m^3-n^3+k^3)f_0\right)}$$

$$\frac{1 - e^{-2\alpha_{n_s} z} e^{j4\pi^2 f_0^2 (m-n)(k-n)\left(\beta_{2,n_s} + \pi \beta_{3,n_s}(m+k)f_0\right)z}}{2\alpha_{n_s} - j4\pi^2 f_0^2 (m-n)(k-n)\left(\beta_{2,n_s} + \pi \beta_{3,n_s}(m+k)f_0\right)}$$

(B23)

At the end of the $n_s$-th span, considering the lumped dispersion and power gain, the NLI field is,

$$E_{NLI}^{(n_s)}\left(\sum_{n'_s=1}^{n_s} L_{s,n'_s}, f\right) =$$

$$= -j\gamma_{n_s} f_0^{\frac{3}{2}} \prod_{n'_s=1}^{n_s-1} g_{n'_s}^{3/2} e^{-3\alpha_{n'_s} L_{s,n'_s}} g_{n_s}^{1/2} e^{-\alpha_{n_s} L_{s,n_s}}$$

$$\sum_{i=-\infty}^{\infty} \delta(f-if_0) e^{-j2\pi^2 i^2 f_0^2 \left(\left(\beta_{2,n_s} + \frac{2}{3}\pi \beta_{3,n_s} if_0\right) L_{s,n_s} + \beta_{DCU,n_s}\right)}$$

$$\sum_{m,n,k \in \tilde{A}_i} \xi_m \xi_n^* \xi_k \sqrt{G_{TX}(mf_0) G_{TX}(nf_0) G_{TX}(kf_0)}$$

$$e^{-j2\pi^2 f_0^2 \sum_{n'_s=1}^{n_s-1}\left(\left(\beta_{2,n'_s} L_{s,n'_s} + \beta_{DCU,n'_s}\right)(m^2-n^2+k^2) + \frac{2}{3}\pi \beta_{3,n'_s} L_{s,n'_s}(m^3-n^3+k^3)f_0\right)}$$

$$\frac{1 - e^{-2\alpha_{n_s} L_{s,n_s}} e^{j4\pi^2 f_0^2 (m-n)(k-n)\left(\beta_{2,n_s} + \pi \beta_{3,n_s}(m+k)f_0\right) L_{s,n_s}}}{2\alpha_{n_s} - j4\pi^2 f_0^2 (m-n)(k-n)\left(\beta_{2,n_s} + \pi \beta_{3,n_s}(m+k)f_0\right)}$$

(B23)

In the end of the link, the NLI field generated by the $n_s$-th span is,

$$E_{NLI}^{(n_s)}(L_{tot},f) = e^{-j2\pi^2(if_0)^2 \sum_{n'_s=n_s+1}^{N_s}\left(\beta_{2,n'_s}L_{s,n'_s}+\frac{2}{3}\pi if_0\beta_{3,n'_s}L_{s,n'_s}+\beta_{DCU,n'_s}\right)}$$

$$\prod_{n'_s=n_s+1}^{N_s} g_{n'_s}^{1/2} e^{-\alpha_{n'_s}L_{s,n'_s}} \cdot E_{NLI}^{(n_s)}\left(\sum_{n'_s=1}^{n_s}L_{s,n'_s},f\right)$$

$$= -j\gamma_{n_s}f_0^{\frac{3}{2}} \prod_{n'_s=1}^{n_s-1} g_{n'_s}^{3/2} e^{-3\alpha_{n'_s}L_{s,n'_s}} \prod_{n'_s=n_s}^{N_s} g_{n'_s}^{1/2} e^{-\alpha_{n'_s}L_{s,n'_s}} \sum_{i=-\infty}^{\infty}\delta(f-if_0)$$

$$e^{-j2\pi^2(if_0)^2 \sum_{n'_s=1}^{N_s}\left(\beta_{2,n'_s}L_{s,n'_s}+\frac{2}{3}\pi if_0\beta_{3,n'_s}L_{s,n'_s}+\beta_{DCU,n'_s}\right)}$$

$$\sum_{m,n,k\in\tilde{A}_i}\xi_m\xi_n^*\xi_k\sqrt{G_{TX}(mf_0)G_{TX}(nf_0)G_{TX}(kf_0)}$$

$$e^{j4\pi^2 f_0^2 (m-n)(k-n)\sum_{n'_s=1}^{n_s-1}\left((\beta_{2,n'_s}L_{s,n'_s}+\beta_{DCU,n'_s})+\pi\beta_{3,n'_s}L_{s,n'_s}(m+k)f_0\right)}$$

$$\frac{1-e^{-2\alpha_{n_s}L_{s,n_s}}e^{j4\pi^2 f_0^2(m-n)(k-n)(\beta_{2,n_s}+\pi\beta_{3,n_s}(m+k)f_0)L_{s,n_s}}}{2\alpha_{n_s}-j4\pi^2 f_0^2(m-n)(k-n)(\beta_{2,n_s}+\pi\beta_{3,n_s}(m+k)f_0)}$$

(B24)

Therefore, the resulting NLI field is the sum of all NLI fields which are generated by the total $N_s$ spans,

$$E_{NLI}(L_{tot},f) = \sum_{n_s=1}^{N_s} E_{NLI}^{(n_s)}(L_{tot},f) \quad (B25)$$

The final result is shown in Eq. (97).